\documentclass{jfm}
\usepackage{graphicx}
\usepackage{epstopdf, epsfig}

\usepackage{tikz}
\usetikzlibrary{intersections}

\graphicspath{ {images/} }

\shorttitle{Dynamics of localized states in plane Couette flow}
\shortauthor{A. Pershin, C. Beaume and S. M.\ Tobias}

\title{Dynamics of spatially localized states in transitional plane Couette flow}

\author{Anton Pershin\aff{1}
  \corresp{\email{mmap@leeds.ac.uk}},
  C\'edric Beaume\aff{1}
 \and Steven M. Tobias\aff{1}}

\affiliation{\aff{1}School of Mathematics, University of Leeds, Leeds LS2 9JT, UK}

\begin{document}

\maketitle

\begin{abstract}
Unsteady spatially localized states such as puffs, slugs or spots play an important role in transition to turbulence. In plane Couette flow, steady versions of these states are found on two intertwined solution branches describing homoclinic snaking \citep{Schneider:2010_snakes_and_ladders}. These branches can be used to generate a number of spatially localized initial conditions whose transition can be investigated. From the low Reynolds numbers where homoclinic snaking is first observed ($\Rey < 175$) to transitional ones ($\Rey \approx 325$), these spatially localized states traverse various regimes where their relaminarisation time and dynamics are affected by the dynamical structure of phase space. These regimes are reported and characterised in this paper for a $4\pi$ periodic domain in the streamwise direction as a function of the two remaining variables: the Reynolds number and the width of the localized pattern. Close to the snaking, localized states are attracted by spatially localized periodic orbits before relaminarising. At larger values of the Reynolds number, the flow enters a chaotic transient of variable duration before relaminarising. Very long chaotic transients ($t > 10^4$) can be observed without difficulty for relatively low values of the Reynolds number ($\Rey \approx 250$).
\end{abstract}

\begin{keywords}
Authors should not enter keywords on the manuscript, as these must be chosen by the author during the online submission process and will then be added during the typesetting process (see http://journals.cambridge.org/data/\linebreak[3]relatedlink/jfm-\linebreak[3]keywords.pdf for the full list)
\end{keywords}


\section{Introduction}

Transition to turbulence is often studied in simple subcritical flows where turbulence is found for a range of parameter values for which the laminar flow is linearly stable \citep{Orszag71,Romanov:1973_Stability_of_pCf,Meseguer03}.
Owing to the subcritical nature of these flows, methods based on weakly nonlinear theory are of no use and alternative methods have to be employed.

Plane Couette flow, the viscous three-dimensional flow between two oppositely moving parallel plates, was experimentally found to transition to turbulence via the spreading of turbulent spots, which resulted in the estimation of a critical Reynolds number for transition: $\Rey_c = 325 \pm 5$ \citep{Dauchot:1995_Finite_ampl_pert_and_spots}.
More recently, intensive numerical investigations of localized turbulence in shear flows using statistical methods were undertaken to locate regime changes.
Important results came in first in pipe flow through the characterisation of turbulent lifetimes \citep{Eckhardt07, Willis:2007_Critical_behaviour_in_relam_of_loc_turb_pipe, Avila:2010_On_transient_nature_of_localised_pipe}.
This effort culminated with the identification of a statistical critical Reynolds number based on the comparison between the mean lifetime of a decaying turbulent puff and the timescale corresponding to its splitting \citep{Avila:2011_onset_of_turbulence}.
The same technique was utilised for plane Couette flow, where the statistical threshold where turbulent stripe splitting and decay lifetimes intersect was determined to be $\Rey_c \approx 325$ \citep{Shi:2013_scale_invariance_at_onset_in_pcf}, in remarkable agreement with the experiment by \citet{Dauchot:1995_Finite_ampl_pert_and_spots}.
The observable bistability between laminar and turbulent flows motivated another type of research based on front propagation \citep{Pomeau1986,Duguet13}.
Recent studies highlight the similarities between transition to turbulence and directed percolation 
\citep{Sano:2016_Universal_trans_to_turb, Lemoult:2016_directed_percolation_in_couette_flow, Chantry:2017_universal_transition_planar_shear_flow}.
Even though these statistical approaches can be used to locate the transition threshold, they only provide a limited understanding of the circumstances under which transition occurs, which is necessary to design control strategies.

Other research has focused on the boundary between the basins of attraction of the laminar and of the turbulent flows known as the edge of chaos \citep{Skufca:2006_edge_of_chaos} both for pipe flow \citep{Schneider:2007_turbulence_transition_and_edge_of_chaos_in_pipe_flow} and plane Couette flow \citep{Schneider:2008_laminar_turbulent_boundary_in_pcf, Duguet:2009_localized_edge_states}. 
In the latter case, it was found that, while the trajectories on the edge in a small periodic domain converge to solutions closely resembling the lower branch of spatially periodic Nagata solutions \citep{Nagata:1990_3d_finite_amplitude_solutions_in_pcf, Schneider:2008_laminar_turbulent_boundary_in_pcf}, those in large domains appear to be spatially localized \citep{Duguet:2009_localized_edge_states, Schneider:2010_localized_edge_states_nucleate_turbulence}.
When the domain is not extended in the streamwise direction, edge states take the form of exact spatially (spanwise) localized solutions comprised of streamwise-oriented streaks and rolls surrounded by laminar flow. These are either equilibria or travelling waves in the streamwise direction depending on their parity.
These states, which have recently been related to maximum transient growth perturbations \citep{Olvera:2017_optimising_energy_growth}, can be continued down in Reynolds number to unveil a so-called homoclinic snaking bifurcation scenario \citep{Schneider:2010_snakes_and_ladders, Gibson:2016_homoclinic_snaking}.
In this scenario, the branches of localized states oscillate in a bounded region in parameter space where each oscillation corresponds to an increase of the localized pattern by two rolls, one on either side of the pattern.
Homoclinic snaking has been thoroughly studied for the Swift--Hohenberg equation (see \citet{Knobloch:2015_spatial_localization} and references therein) but also in a wide variety of physical systems where forcing is balanced by a dissipative mechanism \citep{Woods99,Mercader11,Beaume:2013_convectons_and_secondary_snaking_in_3d_ddc,Lloyd15}.
In particular, in doubly diffusive convection, the analysis of the homoclinic snaking and the stability of its solutions proved helpful to understand the complex regime that arises directly above criticality \citep{Beaume18}.

Homoclinic snaking is often associated with depinning whereby successive wavelength nucleations at the edge of the localized structure result in the propagation of the front connecting the pattern to the quiescent background. The  speed of the front is found to be  proportional to the square-root of the distance to the snaking \citep{Burke:2006_localized_states_in_generalized_she, Knobloch:2015_spatial_localization}.
The depinning instability is particularly interesting in the case of transition to turbulence since it would provide a potential mechanism by which a non-laminar state could invade the domain.
Although depinning has been observed in two-dimensional doubly diffusive convection \citep{Bergeon08}, the study of the same system in three dimensions has revealed the presence of another instability associated with shorter timescales than depinning, preventing it from being observed \citep{Beaume18}.
Using localized turbulent initial conditions in plane Couette flow, \citet{Duguet:2011_stochastic_and_deterministic_front} observed a small depinning region to the right of the snaking region, which disappears in favour of  stochastic evolution.

In this work, we compute exact localized solutions of plane Couette flow and investigate their dynamics when perturbed in Reynolds number to the right of the snaking region, in order to understand better the mechanisms of transition and relaminarisation.
We use a domain with a width twice that utilised by  \citet{Schneider:2010_snakes_and_ladders}, comparable with the domain used by \citet{Duguet:2011_stochastic_and_deterministic_front}.
In the next section, we detail the flow configuration studied, followed by the description of the spatially localized solutions that we used as initial conditions.
The results are reported in Section 4, with details for the dynamical mechanisms observed in our simulations and the various regions found in parameter space.
Section 5 concludes this paper.

\section{Problem setup}

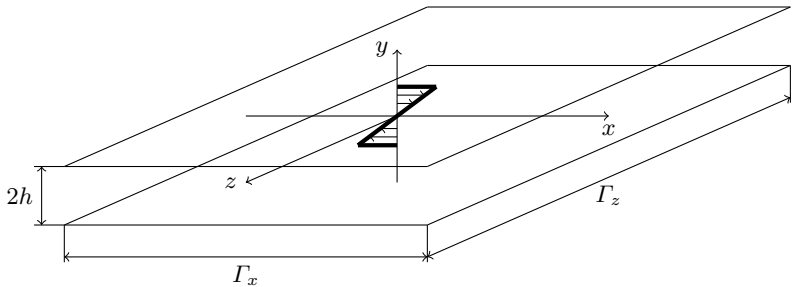
\begin{figure}
  \begin{center}
    \begin{tikzpicture}[y=0.035cm,x=0.4cm]
        \def\ydomainlen{30};
        \def\ydomainslope{5};
        \draw [->] (0,25) -- (12,25) node [below] {$x$};
        \draw [->] (5,0) -- (5,50) node [left] {$y$};
        \draw [->] (5, 25) -- ({-(\ydomainlen-5)/\ydomainslope + 5}, {25 - (\ydomainlen-5)}) node [left] {$z$};
        \def\k{17/2};
        \def\b{-35/2};
        \def\upwall{36};
        \def\downwall{14};
        \def\downdomain{10};
        \def\gap{\upwall - \downwall};
        \draw [ultra thick] ({(\downwall - \b)/(\k)}, \downwall) -- ({(\upwall - \b)/(\k)}, \upwall);
        \draw [ultra thick] (5, \upwall) -- ({(\upwall - \b)/(\k)}, \upwall);
        \draw [ultra thick] (5, \downwall) -- ({(\downwall - \b)/(\k)}, \downwall);
        \foreach \i in {1,2,5,6}
            \draw [->] (5, {\upwall - \i*(\gap)/7}) -- ({(\upwall - \i*(\gap)/7 - \b)/(\k)}, {\upwall - \i*(\gap)/7});
        \draw (12, \upwall) -- ({-\ydomainlen/\ydomainslope + 12}, \upwall - \ydomainlen);
        \draw (12, \upwall) -- ({\ydomainlen/\ydomainslope + 12}, \upwall + \ydomainlen);
        \draw (0, \upwall) -- ({-\ydomainlen/\ydomainslope + 0}, \upwall - \ydomainlen);
        \draw (0, \upwall) -- ({\ydomainlen/\ydomainslope + 0}, \upwall + \ydomainlen);
        \draw ({-\ydomainlen/\ydomainslope + 0}, \upwall - \ydomainlen) -- ({-\ydomainlen/\ydomainslope + 12}, \upwall - \ydomainlen);
        \draw ({\ydomainlen/\ydomainslope + 0}, \upwall + \ydomainlen) -- ({\ydomainlen/\ydomainslope + 12}, \upwall + \ydomainlen);
        \draw (12, \downwall) -- ({-\ydomainlen/\ydomainslope + 12}, \downwall - \ydomainlen);
        \draw (12, \downwall) -- ({\ydomainlen/\ydomainslope + 12}, \downwall + \ydomainlen);
        \draw (0, \downwall) -- ({-\ydomainlen/\ydomainslope + 0}, \downwall - \ydomainlen);
        \draw (0, \downwall) -- ({\ydomainlen/\ydomainslope + 0}, \downwall + \ydomainlen);
        \draw ({-\ydomainlen/\ydomainslope + 0}, \downwall - \ydomainlen) -- ({-\ydomainlen/\ydomainslope + 12}, \downwall - \ydomainlen);
        \draw ({\ydomainlen/\ydomainslope + 0}, \downwall + \ydomainlen) -- ({\ydomainlen/\ydomainslope + 12}, \downwall + \ydomainlen);
        \draw ({-\ydomainlen/\ydomainslope + 0}, \downwall - \ydomainlen) -- ({-\ydomainlen/\ydomainslope + 0 - 1}, \downwall - \ydomainlen);
        \draw ({-\ydomainlen/\ydomainslope + 0}, \upwall - \ydomainlen) -- ({-\ydomainlen/\ydomainslope + 0 - 1}, \upwall - \ydomainlen);
        \draw [<->]({-\ydomainlen/\ydomainslope + 0 - 0.75}, \downwall - \ydomainlen) -- node [left, midway] {$2h$} ({-\ydomainlen/\ydomainslope + 0 - 0.75}, \upwall - \ydomainlen);
        \draw ({-\ydomainlen/\ydomainslope + 0}, \downwall - \ydomainlen) -- ({-\ydomainlen/\ydomainslope + 0}, \downwall - \ydomainlen - 14);
        \draw ({-\ydomainlen/\ydomainslope + 12}, \downwall - \ydomainlen) -- ({-\ydomainlen/\ydomainslope + 12}, \downwall - \ydomainlen - 14);
        \draw [<->]({-\ydomainlen/\ydomainslope + 0}, \downwall - \ydomainlen - 12) -- node [below, midway] {$\Gamma_x$} ({-\ydomainlen/\ydomainslope + 12}, \downwall - \ydomainlen - 12);
        \draw ({\ydomainlen/\ydomainslope + 12}, \downwall + \ydomainlen) -- ({\ydomainlen/\ydomainslope + 12}, \downwall + \ydomainlen - 14);
        \draw [<->]({-\ydomainlen/\ydomainslope + 12}, \downwall - \ydomainlen - 12) -- node [below, midway] {$\Gamma_z$} ({\ydomainlen/\ydomainslope + 12}, \downwall + \ydomainlen - 12);
    \end{tikzpicture}
    \end{center}
    \caption{Sketch of the plane Couette flow configuration and its laminar solution (thick black line). The $y=\pm 1$ walls are parallel, separated by $2h$ and move along the $x$-direction with velocity $\pm U$. The domain is considered periodic in both $x$ and $z$ with period $\Gamma_x$ and $\Gamma_z$ respectively.}
    \label{fig:pcf_layout}
\end{figure}

We consider plane Couette flow, which is a three-dimensional flow confined between two parallel walls moving in opposite directions as shown in figure \ref{fig:pcf_layout}.
The dynamics of this flow is governed by the Navier--Stokes equation:
\begin{equation}
\label{eq:NSE}
\partial_t \boldsymbol{u} + (\boldsymbol{u} \cdot \nabla) \boldsymbol{u} = -\nabla p + \frac{1}{\Rey} \nabla^2 \boldsymbol{u},
\end{equation}
where $\boldsymbol{u}$ is the velocity field with components $u$, $v$ and $w$ in the streamwise ($x$), wall-normal ($y$) and spanwise ($z$) directions respectively, $p$ is the pressure and $t$ is the time.
The Navier--Stokes equation is accompanied with the incompressibility condition:
\begin{equation}
\label{eq:incompr}
\nabla \cdot \boldsymbol{u} = 0.
\end{equation}
These equations have been nondimensionalised using the speed of the walls $U$ and half the gap between them, $h$, as units of velocity and distance.
The Reynolds number in equation (\ref{eq:NSE}) is:
\begin{equation}
\Rey = \frac{U h}{\nu},
\end{equation}
where $\nu$ is the kinematic viscosity of fluid.
We consider  periodic boundary conditions in the streamwise and spanwise directions: $\boldsymbol{u}(x, y, z) = \boldsymbol{u}(x + \Gamma_x, y, z + \Gamma_z)$, where $\Gamma_x = L_x / h$ and $\Gamma_z = L_z / h$ are the nondimensional spatial periodicities in the streamwise and spanwise directions.
No-slip boundary conditions are used in the remaining, wall-normal direction: $\boldsymbol{u}|_{y = \pm1} = (\pm 1, 0, 0)$.

Plane Couette flow possesses a laminar solution: the parallel, unidirectional flow: $\boldsymbol{U} = (y, 0, 0)$ associated with constant pressure.
To characterise the flow, we track the dynamics of the departures $\boldsymbol{\tilde{u}}$ from this state by writing: $\boldsymbol{u} = \boldsymbol{U} + \boldsymbol{\tilde{u}}$ \citep{Schmid:2001_stability_and_transition_in_shear_flows} and thus solve the system:
\begin{subeqnarray}
\partial_t \boldsymbol{\tilde{u}} + \tilde{v} \boldsymbol{e_x} + y \partial_x \boldsymbol{\tilde{u}} + (\boldsymbol{\tilde{u}} \cdot \nabla) \boldsymbol{\tilde{u}} &=& -\nabla \tilde{p} + \frac{1}{\Rey} \nabla^2 \boldsymbol{\tilde{u}}, \label{eq:NSE_flucs} \\[3pt]
\nabla \cdot \boldsymbol{\tilde{u}} &=& 0, \label{eq:INCOMPR_flucs}
\end{subeqnarray}
where $\boldsymbol{e_x}$ is the unit vector in the $x$-direction.
For notational simplicity, we hereafter drop the tildes.

Our numerical simulations are carried out using \textit{Channelflow} which relies upon a Fourier--Chebyshev--Fourier spatial discretisation and provides a variety of temporal schemes as well as a set of tools for numerical continuation and stability analysis \citep{Gibson:2014_channelflow}.
Localized states in the spanwise direction are the simplest family of spatially localized states in plane Couette flow, so we restrict ourselves to large spanwise domains: $\Gamma_x \times \Gamma_y \times \Gamma_z = 4 \pi \times 2 \times 32 \pi$.
The domain is meshed using $N_x=32$ and $N_z=512$ Fourier points in the streamwise and spanwise directions and $N_y=33$ Chebyshev points in the wall-normal direction, a similar discretisation to that found in the literature \citep{Schneider:2010_snakes_and_ladders}.
To time integrate, we use \textit{Channelflow}'s 3rd-order semi-implicit backward differentiation with time step $\triangle t = 1 / \Rey$.
All the time integration results shown here have been obtained without the imposition of any symmetry.

\section{Initial conditions}

\begin{figure}
    \begin{center}
    \includegraphics[width=1\textwidth]{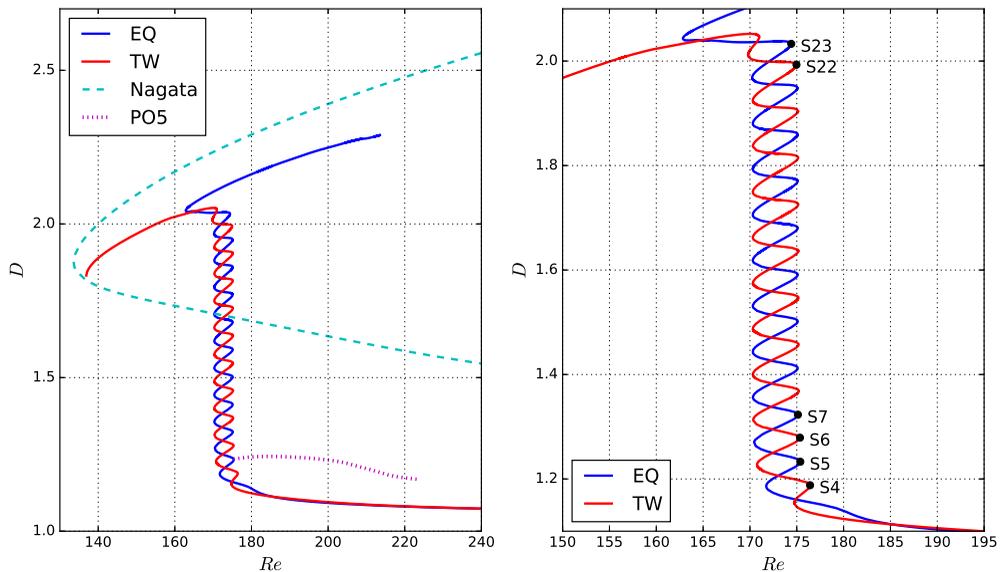}
    \end{center}
    \caption{Bifurcation diagram of the homoclinic snaking described by the localized equilibria EQ and travelling waves TW (blue and red lines respectively) of plane Couette flow. The solutions are represented using their enstrophy $D$ (defined in the text) plotted against the Reynolds number $\Rey$. The Nagata solutions (dashed line) and the localized periodic orbits PO5 (dotted line) are shown on the left panel (PO5 will be discussed in the following sections). The right panel is an enlargement of the snakes that identifies some of the initial conditions (S4, S5, S6, S7, S22 and S23) used in this paper.}
    \label{fig:snaking}
\end{figure}

\begin{figure}
    \begin{center}
    \includegraphics[width=1\textwidth]{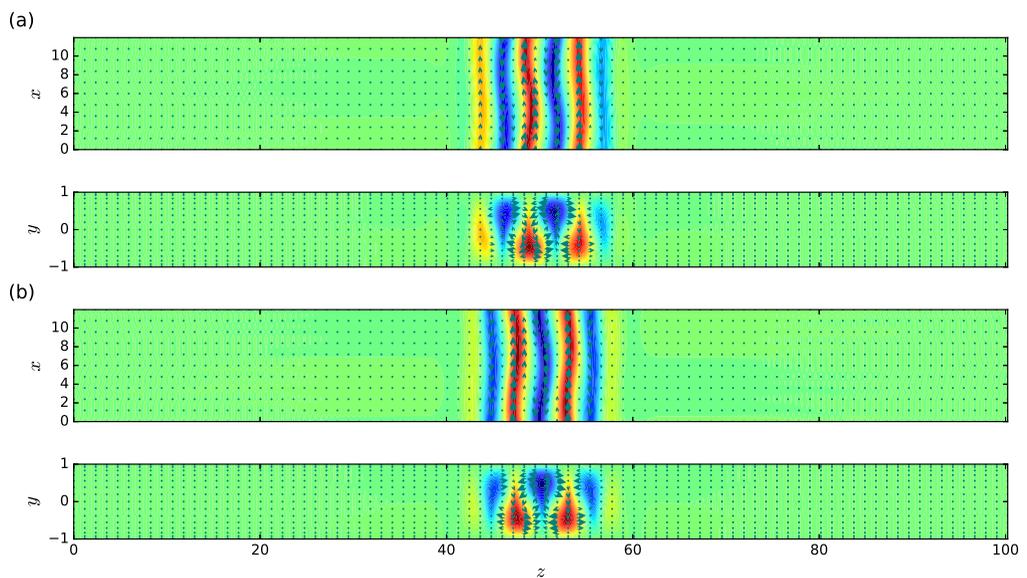}
    \end{center}
    \caption{Localized (a) equilibrium lying on the lower branch of EQ at $\Rey \approx 200.055$ and (b) travelling wave lying on the lower branch of TW at $\Rey \approx 202.494$. Solutions are represented using the streamwise velocity in colour and the in-plane velocity with arrows on the $y$-averaged plane (top panels) and $x$-averaged plane (bottom panels).}
    \label{fig:fields_low_branches}
\end{figure}

There are two families of simple spatially localized states in plane Couette flow: equilibria (hereafter EQ), which are symmetric with respect to the reflection: $[u, v, w](x, y, z) \longrightarrow -[u, v, w](-x, -y, -z)$, and streamwise travelling waves (hereafter TW) which are shift-reflect-symmetric ($[u, v, w](x, y, z) \longrightarrow [u, v, -w](x + \Gamma_x / 2, y, -z)$).
These two solutions define branches on the bifurcation diagram in figure \ref{fig:snaking} that exhibit homoclinic snaking \citep{Woods99,Burke:2006_localized_states_in_generalized_she,Schneider:2010_snakes_and_ladders,Knobloch:2015_spatial_localization}.
On this figure, solutions are characterised using the enstrophy:
\begin{equation}
D = \frac{1}{\Gamma_x \Gamma_y \Gamma_z} \int \int \int_{\Omega} |\nabla \times \boldsymbol{u}|^2 dx dy dz,
\end{equation}
where $\Omega$ is the computational domain and where $D = 1$ for the laminar solution.
Localized states dissipate more energy than the laminar solution ($D>1$) and, for large enough $Re$, are comprised of 3 (resp. 2) streamwise-oriented rolls for EQ (resp. TW) as shown in figure \ref{fig:fields_low_branches}. 
As the branches undergo one back-and-forth oscillation in Reynolds number in the direction of increasing energy, the size of structure is increased by the nucleation of one roll on each side of the localized structure. 
As a result, the localized states grow as one moves up the snakes while still preserving their symmetry and parity: EQ always displays an odd number of rolls while TW always displays an even number of rolls.
At the top of the snakes, the solutions are domain-filling and TW continues to lower values of $\Rey$ to reconnect the unstable spatially periodic Nagata solution \citep{Gibson:2009_equilibrium_and_tw_in_pcf} close to its saddle-node (see figure \ref{fig:snaking}).
The branch EQ differs in behaviour: it instead moves to higher values of $\Rey$ while the solution resembles the Nagata solution except for a defect due to the lack of available room for the state to nucleate an additional roll and fill the domain \citep{Bergeon08}.
At the bottom of the branch EQ, a subcritical Hopf bifurcation occuring slightly above the lowest saddle-node, gives rise to a branch of spatially localized reflection-symmetric periodic orbits, namely PO5. 
Oscillatory instabilities near the lowest saddle-node of a snaking branch were predicted by \citet{Burke:2012_localized_states_in_extended_she} 
in non-variational and non-conservative systems.

We select the set comprised of the right saddle-node states of both snakes as our initial conditions for time integration.
For clarity, we name these according to their roll count: the right saddle-node states of EQ are thus named S5, S7, etc. and those of TW are named S4, S6, etc.
Note that all the localized states computed here are unstable \citep{Schneider:2010_localized_edge_states_nucleate_turbulence} and that the larger the localized pattern, the more unstable it is \citep{Gibson:2016_homoclinic_snaking}.
This is the case in another three-dimensional symmetric fluid system \citep{Beaume18}.

In order to compare the relaminaristation dynamics of each initial condition at various values of the Reynolds number, we calculate the associated relaminarisation time $t_{relam}$, i.e. the time it takes for the flow to reach a small (attracting) neighbourhood of the laminar fixed point where the flow dynamics is well-described by the linearised Navier--Stokes equation. 
We denote the time-dependent maximum pointwise kinetic energy $E_{max}(t)~= \max\limits_{\mathbf{x}}{\frac{1}{2}|\mathbf{u}(\mathbf{x}, t)|^2}$ and seek $t_{relam}$ such that $E_{max}$ decays exponentially for $t > t_{relam}$.
To set a similar condition on $E_{max}$, we ran a number of preliminary simulations which all displayed chaotic oscillations around $O(1)$ values before  relaminarising.
We observed that, for $E_{max} < 0.1$, all our simulations decayed exponentially.
Even though the basin of attraction of the laminar fixed point is wider than the mere region of exponential decay of $E_{max}(t)$, we acted out of caution and defined $t_{relam}$ by solving  $E_{max}(t_{relam}) = 0.1$.

\section{Main results}

We take the states Si, with i=$4, 5, \dots, 23$, as initial conditions and time-integrate for a range of Reynolds numbers $\Rey \in (\Rey_s(\textrm{i}); 350]$ where $\Rey_s(\textrm{i})$ is the Reynolds number associated with Si, as shown in table \ref{sntable}. 
\begin{table}
\begin{center}
\def~{\hphantom{0}}
\begin{tabular}{lccccccccccc}
Initial condition & S4 & S5 & S6 & S7 & S8 & S9 & S10 & S11 \\
Reynolds number & $176.423$ & $175.375$ & $175.347$ & $175.124$ & $175.229$ & $175.097$ & $175.199$ & $175.105$ \\
~&~&~&~&~&~&~&~&~&~&~&~\\
Initial condition & S12 & S13 & S14 & S15 & S16 & S17 & S18 & S19 \\
Reynolds number & $175.180$ & $175.104$ & $175.165$ & $175.105$ & $175.161$ & $175.104$ & $175.150$ & $175.105$ \\
~&~&~&~&~&~&~&~&~&~&~&~\\
Initial condition & S20 & S21 & S22 & S23 \\
Reynolds number & $175.142$ & $175.088$ & $174.971$ & $174.414$
\end{tabular}
\caption{Reynolds number corresponding to the various initial conditions considered.}
\label{sntable}
\end{center}
\end{table}
Before embarking on the mechanism for relaminarisation, we shall discuss the typical dynamics undergone by the flow.

\subsection{Flow dynamics}\label{cyclic}

\begin{figure}
    \begin{center}
    \includegraphics[width=1\textwidth]{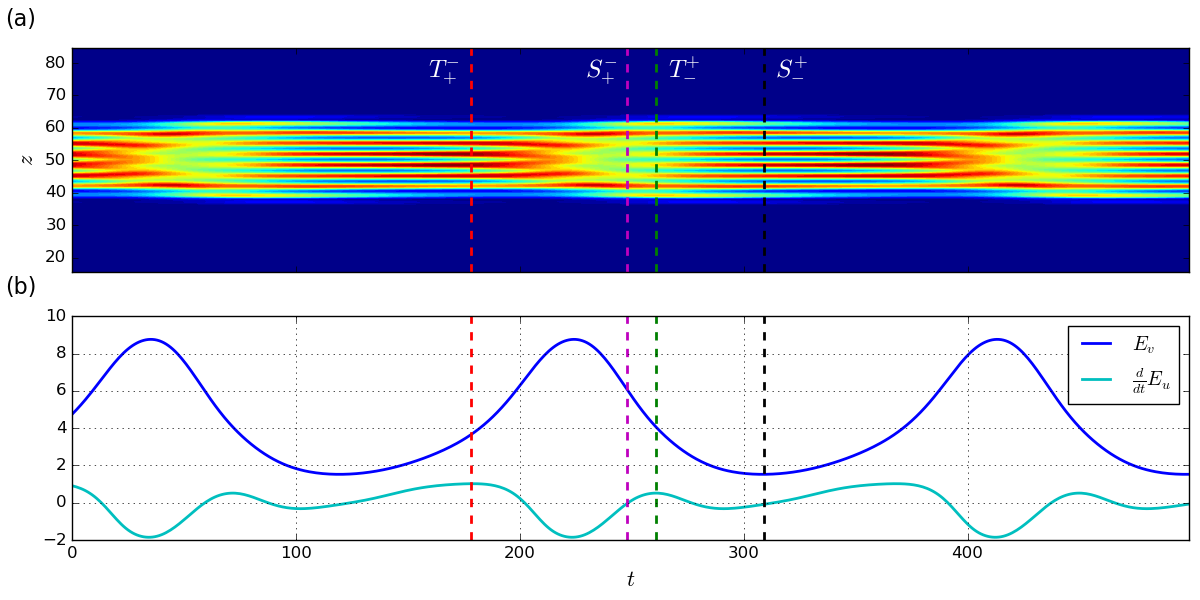}
    \end{center}
    \caption{Dynamics of PO5 at $\Rey = 200.41072$ represented through a space-time plot of the streamwise- and wall-normal-averaged kinetic energy $E_{xy}$ (a) and through the temporal evolution of the roll kinetic energy $E_{v}$ (dark blue) and of the growth rate associated with the streaks kinetic energy $E_{u}$ (light blue) (b). 
    The panels are aligned so that they share the temporal scale ($x$-axis). All quantities are defined in the text and the time is arbitrarily set to $0$ at the beginning of the displayed time-interval. Vertical dashed lines correspond to special points as shown in figure \ref{fig:selected_corr}.
    On the space-time plot, blue region corresponds to the laminar flow ($E_{xy} = 0$) and red-colored regions correspond to the largest values of $E_{xy}$.}
    \label{fig:po_ke_and_modes}
\end{figure}

\begin{figure}
    \begin{center}
    \includegraphics[width=1\textwidth]{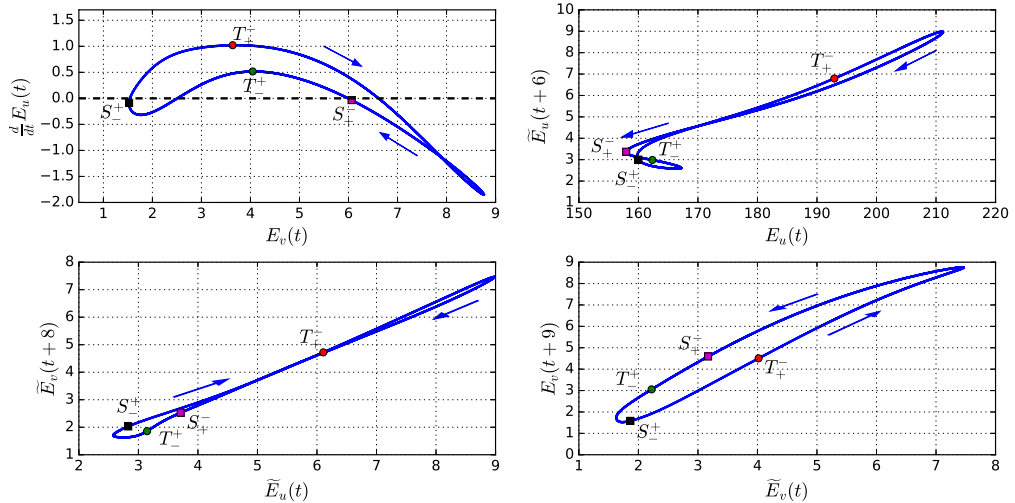}
    \end{center}
    \caption{Correlations between $E_{u}$, $\widetilde{E}_{u}$, $E_{v}$ and $\widetilde{E}_{v}$. The quantities on the $y$-coordinates are either time-derivatives or shifted forward in time. The labelled circles and squares correspond to special points discussed in the text. The phase at these points is indicated in figure \ref{fig:po_ke_and_modes} using vertical dashed lines of the corresponding colour.}
    \label{fig:selected_corr}
\end{figure}

The flows studied are of oscillatory nature, with oscillations typically replicating the dynamics found on the periodic orbit PO5.
To illustrate the flow dynamics, we consider a localized periodic orbit with pattern wavelength $l_z \approx 6.7$ taken from PO5 at $\Rey = 200.41072$ and shown in figure \ref{fig:po_ke_and_modes}.
To characterise it, we use the streamwise- and wall-normal-averaged kinetic energy:
\begin{equation}
E_{xy}(z) = \frac{1}{2}\int \limits_{0}^{\Gamma_x} \int \limits_{0}^{\Gamma_y} |\boldsymbol{u}|^2 dx dy.
\end{equation}
Figure \ref{fig:po_ke_and_modes}(a) shows typical oscillations described by the flow, dominated by streaks whose amplitude oscillates in time while a nucleation event starts but never reaches completion.
For the parameter values used, the oscillations have period $T \approx 189$.
The period of oscillation is a function of the Reynolds number but the description below remains qualitatively accurate for most of the values of the Reynolds number studied.
To gain more insight, we decompose the flow field in Fourier series in the streamwise direction \citep{Wang:2007_lower_branch}:
\begin{equation}
\boldsymbol{u} = \sum_{k=0}^{\infty} \boldsymbol{u}_k(y, z) e^{i \alpha k x} + c.c.,
\end{equation}
where $\alpha = 2 \pi / \Gamma_x$, $k$ is a positive integer, and $c.c.$ stands for the complex conjugate expression.
This decomposition allows the introduction of the following quantities:
\begin{subeqnarray}
E_{u} &=& \frac{1}{\Gamma_y l_z} \int_{0}^{\Gamma_y} \int_{\Gamma_z / 2}^{\Gamma_z / 2 + l_z} u_0(y, z)^2 dy dz, \\[3pt]
\widetilde{E}_{u} &=& \frac{1}{\Gamma_y l_z} \int_{0}^{\Gamma_y} \int_{\Gamma_z / 2}^{\Gamma_z / 2 + l_z} u_1(y, z)^2 dy dz, \\[3pt]
E_{v} &=& \frac{1}{\Gamma_y l_z} \int_{0}^{\Gamma_y} \int_{\Gamma_z / 2}^{\Gamma_z / 2 + l_z} \left[v_0(y, z)^2 + w_0(y, z)^2 \right] dy dz, \\[3pt]
\widetilde{E}_{v} &=& \frac{1}{\Gamma_y l_z} \int_{0}^{\Gamma_y} \int_{\Gamma_z / 2}^{\Gamma_z / 2 + l_z} \left[v_1(y, z)^2 + w_1(y, z)^2 \right] dy dz,
\end{subeqnarray}
where we have integrated over one spanwise wavelength located in the centre of the localized structure to minimise the influence of the fronts.
These quantities relate to the internal dynamics of the state: $E_u$ is associated to the streak kinetic energy, $E_v$ to that of the rolls while $\widetilde{E}_{u}$ and $\widetilde{E}_{v}$ are associated to those of the streak and roll fluctuations respectively.
To unravel dynamical mechanisms, we studied the correlations between $E_{u}, \widetilde{E}_{u}, E_{v}, \widetilde{E}_{v}$ and their time-derivatives and report the most useful results in figure \ref{fig:selected_corr}.
The solution displays low amplitude rolls at $t \approx 120$ in figure \ref{fig:po_ke_and_modes} corresponding to the point $S_-^+$ in figure \ref{fig:selected_corr}.
For roll energies $E_v \lessapprox 3.63$, the growth rate of the streaks $dE_u/dt$ is positively correlated with the roll energy and positive, leading to streak growth (top left panel of fig \ref{fig:selected_corr}).
The growth of the streaks leads to the growth of the streak fluctuations, approximately $6$ time units later, as indicated in the top right panel of figure \ref{fig:selected_corr}.
This step is followed, again with a time lag of about $8$ units, by the growth of the roll fluctuations (bottom left panel of figure \ref{fig:selected_corr}) and then, $9$ units of time later, by that of the rolls (bottom right panel of figure \ref{fig:selected_corr}).
This overall growth of the pattern continues beyond the turning point $T_+^-$ at $E_v \approx 3.63$, where the roll energy and the streak growth become anti-correlated.
The shift in dynamics past this turning point is illustrated in figure figure \ref{fig:po_ke_and_modes}(b). The reciprocal shift, where roll energy and streak growth become positively correlated, occurs at $T_-^+$ when $E_v \approx 4.05$.
When the roll energy reaches $E_v \approx 6.61$, the roll overgrowth leads to the decay of the streaks and, successively, to the delayed decay of the streak fluctuations, roll fluctuations and eventually rolls, as shown by all the positive correlations in figure \ref{fig:selected_corr}.
Before closing the loop, the flow readjusts at low amplitude through an event where the streaks grow and then decay again between $S_+^-$ and $S_-^+$ and have a negative correlation with their fluctuations.
These cycles orchestrate the dynamics within the localized pattern at most of the Reynolds numbers we investigated and only quantitative changes of the energy thresholds have been noticed, without any impact on the qualitative behaviour analysed in this section.

\begin{figure}
    \begin{center}
    \includegraphics[width=1\textwidth]{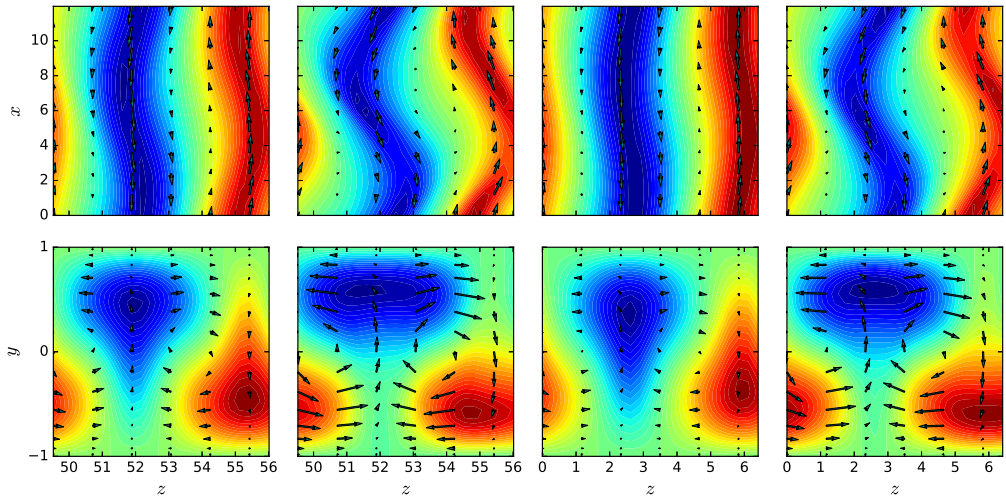}
    \end{center}
    \caption{Snapshots along the cycle at $t \approx 300$ (first column) and $t \approx 410$ (second column) for $\Rey = 200.41072$ accompanied with the lower (third column) and upper (fourth column) Nagata branch states at $\Rey \approx 202.3$.}
    \label{fig:po_and_nag}
\end{figure}

Figure \ref{fig:po_and_nag} shows the velocity field along the cycle at $t \approx 300$, as $E_v$ reaches its minimum and at $t \approx 410$ as $E_v$ reaches its maximum for $Re = 200.41072$ and compares it to the Nagata solutions (lower and upper branch states) at a similar value of the parameter.
When the rolls are at their weakest during the cycle, the flow closely resembles the lower branch Nagata solution and as the rolls are at their strongest, the flow looks similar to the upper branch Nagata solution.
As such, this cyclic dynamics can be seen as the signature of a heteroclinic connection between the two Nagata solutions: the manifold connecting the lower to the upper branch is associated with the reenergising of the streaks by the rolls (together with the delayed growth of the other quantities) and the manifold connecting the upper to the lower branch is associated with the suppression of the streaks due to overgrown rolls.

\subsection{Parameter space map}

\begin{figure}
    \begin{center}
    \includegraphics[width=1\textwidth]{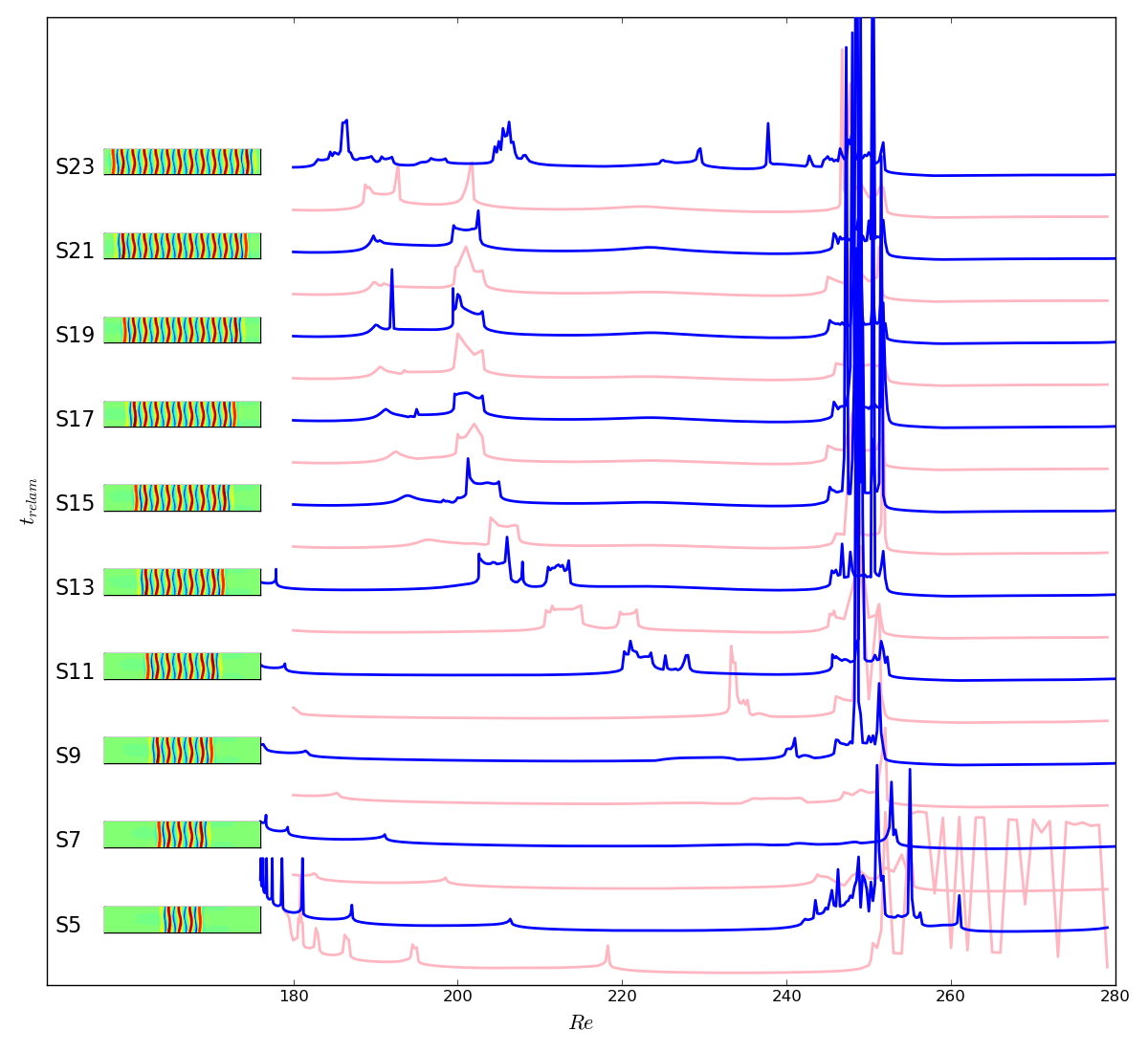}
    \end{center}
    \caption{Relaminarisation times for EQ initial conditions (blue) and TW initial conditions (pink) at $\Rey \in [180; 280]$. The top view of the streamwise velocity of each EQ initial condition is shown on the left. For clarity, each relaminarisation time curve has been displaced upwards by 1000 time units compared to the previous one: $t_{relam} \longleftarrow t_{relam} + 1000 ($i$ - 4)$ for initial condition Si. Corresponding profiles for the initial conditions used (S5--S23) are shown on the left, next to the corresponding curve. As a reference, the plateaux observed on all the curves for $Re < 240$ correspond to $t_{relam} \sim 400$.}
    \label{fig:relam_180_320}
\end{figure}

Figure \ref{fig:relam_180_320} shows the relaminarisation time $t_{relam}$ as a function of $\Rey$ for all our initial conditions. 
We used a discrete sequence of $\Rey \in (\Rey_s; 280]$ whose spacing was manually adjusted and refined in more sensitive regions. 
Our results unveil a complex scenario with a very clear macroscopic organisation in which the size of the initial condition bears more importance than its type, as shown in figure \ref{fig:relam_180_320} where the initial conditions are sorted by the number of rolls in the localized state, thereby alternating EQ and TW initial conditions.
We thus concentrate on EQ only.

\begin{figure}
    \begin{center}
    \includegraphics[width=1\textwidth]{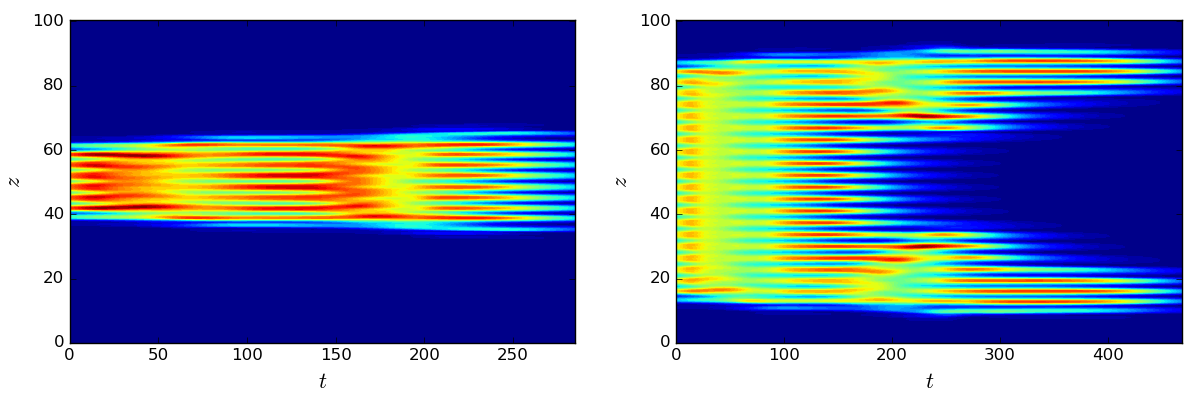}
    \end{center}
    \caption{Dynamics observed on the plateux P and exemplified by the spatiotemporal evolution of the averaged kinetic energy $E_{xy}$ for S5 (left panel) and S19 (right panel) time-integrated at $\Rey = 220$.}
    \label{fig:p_s5_s19_re_220_ke}
\end{figure}

Figure \ref{fig:relam_180_320} also highlights the presence of plateaux of nearly constant relaminarisation times around $t_{relam} = 400$ interspersed with relatively smaller regions of longer-lived transients.
The presence of these plateaux and regions does not appear to be a function of the size of the localized initial conditions, however, the location of these regions is a monotonic function of this quantity.
At low Reynolds numbers, a series of peaks exists in the vicinity of $Re_s$ where the relaminarisation time tends to infinity owing to the fact that the initial condition is a fixed point at $Re_s$.
We shall denote the corresponding region of accumulating peaks as R1. 
Increasing in Reynolds numbers, we refer to the next region of increased relaminarisation time as R2.
The boundaries of R2 move to larger values of the Reynolds number and get closer to each other as the size of the localized pattern decreases and collide between initial conditions S8 and S9.
At even higher Reynolds numbers, between $\Rey \approx 245$ and $\Rey \approx 252$, one encounters another region with long-lived transients: R3.
This region is not sensitive to the size of the initial condition and only exists for Si, i$\ge 8$.
For $\Rey > 280$, most simulations are long-lasting.
We denote this region R4.
The plateaux, surrounding two of the above regions and denoted P, are the locations of the shortest relaminarisation times.
There, the flow does not relaminarise immediately but undergoes oscillations as shown in figure \ref{fig:p_s5_s19_re_220_ke}.
On these plateaux, the localized states typically undergo one or two oscillations before decaying and the wider initial conditions split, giving rise to two spots which, in turn, oscillate before decaying.
This splitting event is a crucial phenomenon for regions R2 and R3 and will be discussed in Sections \ref{rr2} and \ref{rr3}.

The localized initial conditions at the top and bottom of the snaking do not conform to the above general scenario.
The smallest initial conditions develop a different kind of behaviour above $Re = 240$.
In particular, S4 enters R4 at $\Rey \approx 250$ whereas S5 exhibits a region of longer transient between $Re = 242$ and $Re = 261.25$. 
The initial condition S6 displays a similar region to S5, albeit of smaller extent. This region is referred to as R3a.
The wider initial condition S7 displays little dynamic change for $\Rey \gtrsim 190$ except for the presence of one peak at $\Rey \approx 253$.
This peak is a signature of the region R3 which is broader for S8 and reaches its full size for S9.
The largest initial condition, S23, does not follow the above picture either as it is a near domain-filling state.

\begin{figure}
    \begin{tikzpicture}[x=0.86cm, y=0.13cm]
        \def\Remin{0}
        \def\Remax{12}
        \def\ReR{5}
        \def\ReRmid{0.05}
        \def\ReRRbegin{7}
        \def\ReRRend{7.5}
        \def\ReRRmaxbegin{1}
        \def\ReRRmaxend{3.5}
        \def\ReRRRbegin{10.0}
        \def\ReRRRmaxbegin{7.7}
        \def\ReRRRmaxend{9.7}
        \def\ReRRRmidbegin{7.75}
        \def\ReRRRmidend{9.7}
        \def\ReRRRRbegin{9.4}
        \def\ReRRRRmidbegin{11.7}
        \def\ReRRRRmaxbegin{11.8}
        \def\ReRRRRend{14}
        \def\ReRSFIVEbegin{\ReRRRRbegin}
        \def\ReRSFIVEmidbegin{\ReRRRRbegin - 3.3}
        \def\ReRSFIVEmidend{\ReRRRRbegin + 1.5}
        \def\ReRSFIVEmaxbegin{\ReRRRbegin}
        \def\Smin{0}
        \def\Smax{50}
        \def\Smid{20}
        \def\Sfour{0}
        \def\Sfive{4}
        \def\Ssix{8}
        \def\Sseven{12}
        \def\Seight{16}
        \def\Snine{20}
        \def\SRRmin{\Snine}
        \def\SRRRmin{\Sseven}

        \draw (\Remin, \Smin) rectangle (\ReRRRRend, \Smax);
        \draw [->] (\Remin + 5, \Smin - 6) -- (\Remax - 5, \Smin - 6) node [midway, below] (ReText) {$\Rey$};
        \draw [->] (\Remin - 0.9, \Smin + 10) -- (\Remin - 0.9, \Smax - 10) node [midway, above, sloped] (SText) {Saddle-nodes};

        \draw [dashed] (1pt, \Smin) -- (-1pt, \Smin) node [left] {S4};
        \draw [dashed] (\ReRSFIVEmidbegin + 0.9, \Sfive) -- (-1pt, \Sfive) node [left] {S5};
        \draw [dashed] (\ReRRRbegin, \SRRRmin) -- (-1pt, \SRRRmin) node [left] {S7};
        \draw [dashed] (\ReRRbegin, \SRRmin) -- (-1pt, \SRRmin) node [left] {S9};
        \draw (\ReR, 1pt) -- (\ReR, -1pt) node [below] {206};
        \draw [dashed] (\ReRRmaxbegin, \Smax) -- (\ReRRmaxbegin, -1pt) node [below] {190};
        \draw [dashed] (\ReRRmaxend, \Smax) -- (\ReRRmaxend, -1pt) node [below] {203};
        \draw [dashed] (\ReRRRmaxbegin, \Smax) -- (\ReRRRmaxbegin, -1pt) node [below] {245};
        \draw [dashed] (\ReRRRmaxend, \Smax) -- (\ReRRRmaxend, -1pt) node [below] {252};
        \draw [dashed] (\ReRRRRmaxbegin, \Smax) -- (\ReRRRRmaxbegin, -1pt) node [below] {290};
        
        \draw[ultra thick] (\ReR, \Smin) .. controls (2*\ReRmid, \SRRmin - 7) and (\ReRmid, \Smax - 10) .. (\Remin, \Smax);
        \node at (\ReRmid + 1.6, \Sfive + 3) {\textbf{\Large R1}};
        
        \draw[ultra thick] (\ReRRbegin, \SRRmin) .. controls (\ReRRmaxbegin + 0.5, \SRRmin + 10) and (\ReRRmaxbegin + 0.1, \Smax - 10) .. (\ReRRmaxbegin, \Smax);
        \draw[ultra thick] (\ReRRend, \SRRmin) .. controls (\ReRRmaxend - 0.5, \SRRmin + 10) and (\ReRRmaxend + 0.1, \Smax - 10) .. (\ReRRmaxend, \Smax);
        \draw[ultra thick] (\ReRRbegin, \SRRmin) -- (\ReRRend, \SRRmin);
        \node at (\ReRRmaxbegin + 1.4, \Smax - 8) {\begin{tabular}{c} \textbf{\Large R2} \\ \small $t \sim O(10^3)$ \end{tabular}};

        \draw[ultra thick] (\ReRRRbegin, \SRRRmin) .. controls (\ReRRRmidbegin, \SRRRmin + 3) .. (\ReRRRmaxbegin, \Smax);
        \draw[ultra thick] (\ReRRRbegin, \SRRRmin) .. controls (\ReRRRmidend, \SRRRmin + 3) .. (\ReRRRmaxend, \Smax);
        \node at (\ReRRRmaxbegin/2 + \ReRRRmaxend/2, \SRRmin/2 + \Smax/2) {\begin{tabular}{c} \textbf{\Large R3} \\ \small $t \sim O(10^3)$ \end{tabular}};

        \node at (\ReRmid + 2.2, \SRRmin + 5) {\begin{tabular}{c} \textbf{\Large P} \\ \small $t \approx 400$ \end{tabular}};
        \node at (\ReRRRmaxbegin/2 + \ReRRmaxend/2 + 0.4, \SRRmin/2 + \Smax/2 - 3) {\begin{tabular}{c} \textbf{\Large P} \\ \small $t \approx 400$ \end{tabular}};
        \node at (\ReRRRmaxend/2 + \ReRRRRmaxbegin/2, \SRRmin/2 + \Smax/2 - 3) {\begin{tabular}{c} \textbf{\Large P} \\ \small $t \approx 300$ \end{tabular}};
        
        
        \draw[ultra thick] (\ReRSFIVEbegin, \Smin) .. controls (\ReRSFIVEmidbegin, \Sfive - 1.5) and (\ReRSFIVEmidbegin, \Sfive + 1.5) .. (\ReRSFIVEmaxbegin, \SRRRmin);
        \draw[ultra thick] (\ReRSFIVEbegin, \Smin) .. controls (\ReRSFIVEmidend, \Sfive) .. (\ReRSFIVEmaxbegin, \SRRRmin);
        \node at (\ReRRRmaxbegin/2 + \ReRRRmaxend/2 + 0.1, \Sfive + 0.8) {\begin{tabular}{c} \textbf{\Large R3a} \\ \small $t \sim O(10^3)$ \end{tabular}};
        
        \draw[ultra thick] (\ReRRRRbegin, \Smin) .. controls (\ReRRRRmidbegin, \Sfive) .. (\ReRRRRmaxbegin, \Smax);
        \node at (\ReRRRRmaxbegin/2 + \ReRRRRend/2, \Smin/2 + \Smax/2) {\begin{tabular}{c} \textbf{\Large R4} \\ \small $t \gg 10^3$ \end{tabular}};
    \end{tikzpicture}
    \caption{Schematic of the parameter space investigated. Regions R1, R2, R3 and R4 are further studied in Section \ref{rr1} and \ref{rr2} below and are interspersed with plateaux P. In most of these cases, an estimate of the typical observed relaminarisation time is given. Note the variable scale of the horizontal axis, used for readability.}
    \label{fig:regions_schema}
\end{figure}
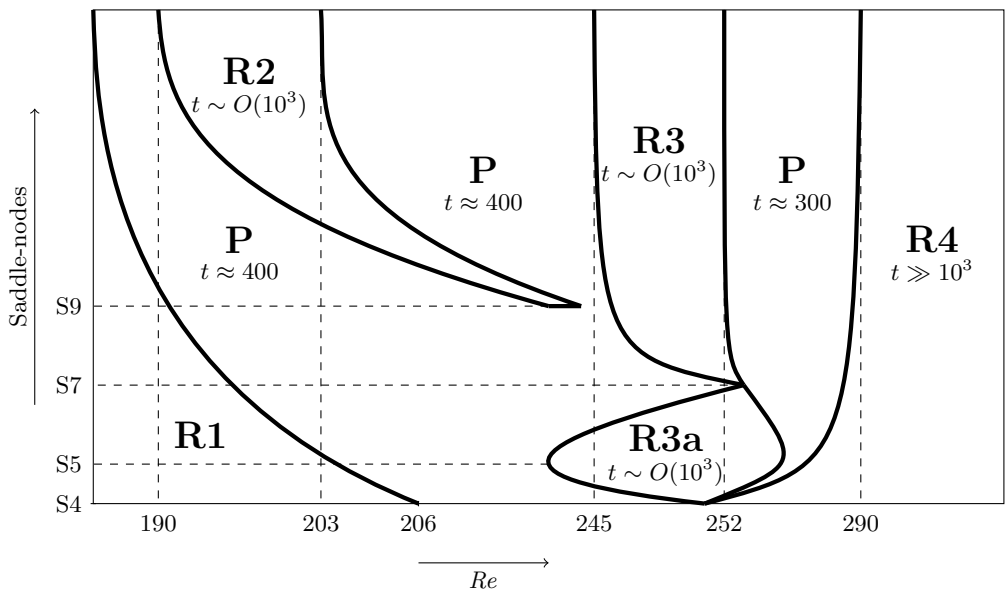

The above observations can be used to provide a schematic of parameter space shown in figure \ref{fig:regions_schema}.
In the next sections, we shall characterise the various regions identified above.

\subsection{Dynamics in the vicinity of the snaking}
\label{rr1}

In the vicinity of the snaking, our initial conditions do not display depinning but rather a well-structured behaviour within region R1 characterised by a series of peaks accumulating at $\Rey_s$ and corresponding to sudden increases of the relaminarisation time as a function of $Re$.
As the width of the initial condition increases, this region becomes smaller and the density of peaks increases.
As this behaviour is similar for all initial conditions, we focus our attention on the two localized EQ initial conditions displaying the largest R1 regions: S5 and S7.

\begin{figure}
    \begin{center}
    \includegraphics[width=1\textwidth]{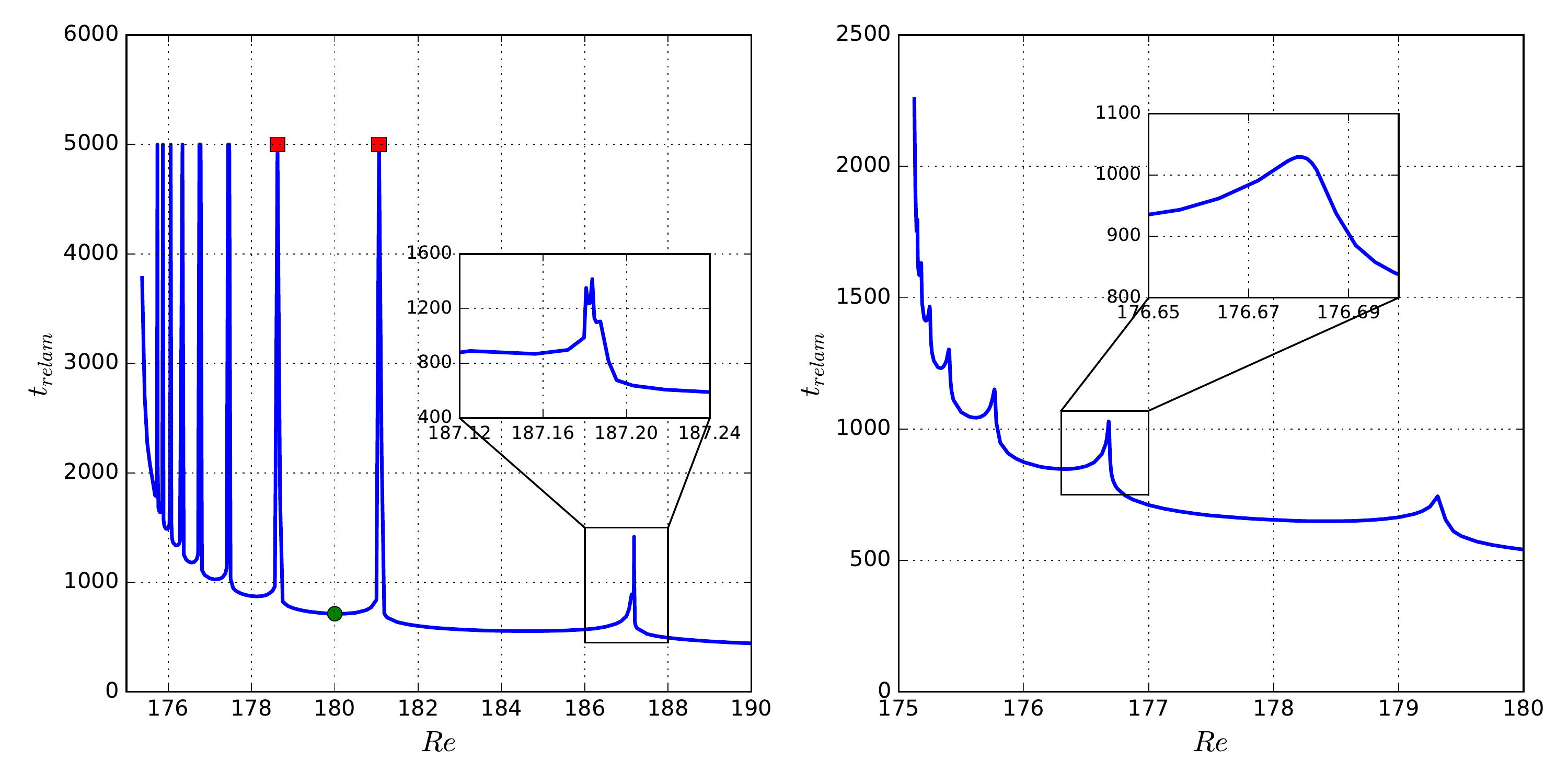}
    \end{center}
    \caption{Enlargements of figure \ref{fig:relam_180_320} for region R1 of initial conditions S5 (left) and S7 (right). The circle denotes the local minimum of $t_{relam}$ located in the region between two neighbouring peaks, marked by squares. All peaks shown in the left plot, except for the one shown in the inset, are associated with infinite relaminarisation times but are here represented with $t_{relam} = 5000$ for clarity. Note that we excluded the rightmost peak of the region, located at $\Rey \approx 208$, for clarity as well. Unlike the left plot, the peaks shown in the right panel have finite $t_{relam}$. }
    \label{fig:r1_s5_and_s7_relam_time}
\end{figure}

The left panel of figure \ref{fig:r1_s5_and_s7_relam_time} shows an enlargement of R1 for S5 where all but the rightmost peak are represented.
Since the initial condition is a (stationary) solution at $\Rey = \Rey_s(5)$, the relaminarisation time tends to infinity at this value.
As we move away from $Re_s(5)$, the relaminarisation time decreases but does not do so monotonically.
Instead, it undergoes a succession of peaks where the relaminarisation time becomes infinite and that accumulates at $Re_s(5)$. 
The last two peaks of S5 are somewhat different.
These peaks, located at $Re \approx 187$ and at $Re \approx 208$ (not shown), display intricate variations and may not lead to infinite timescales due to the more complex dynamics at these higher values of the Reynolds number.

\begin{figure}
    \begin{center}
    \includegraphics[width=1\textwidth]{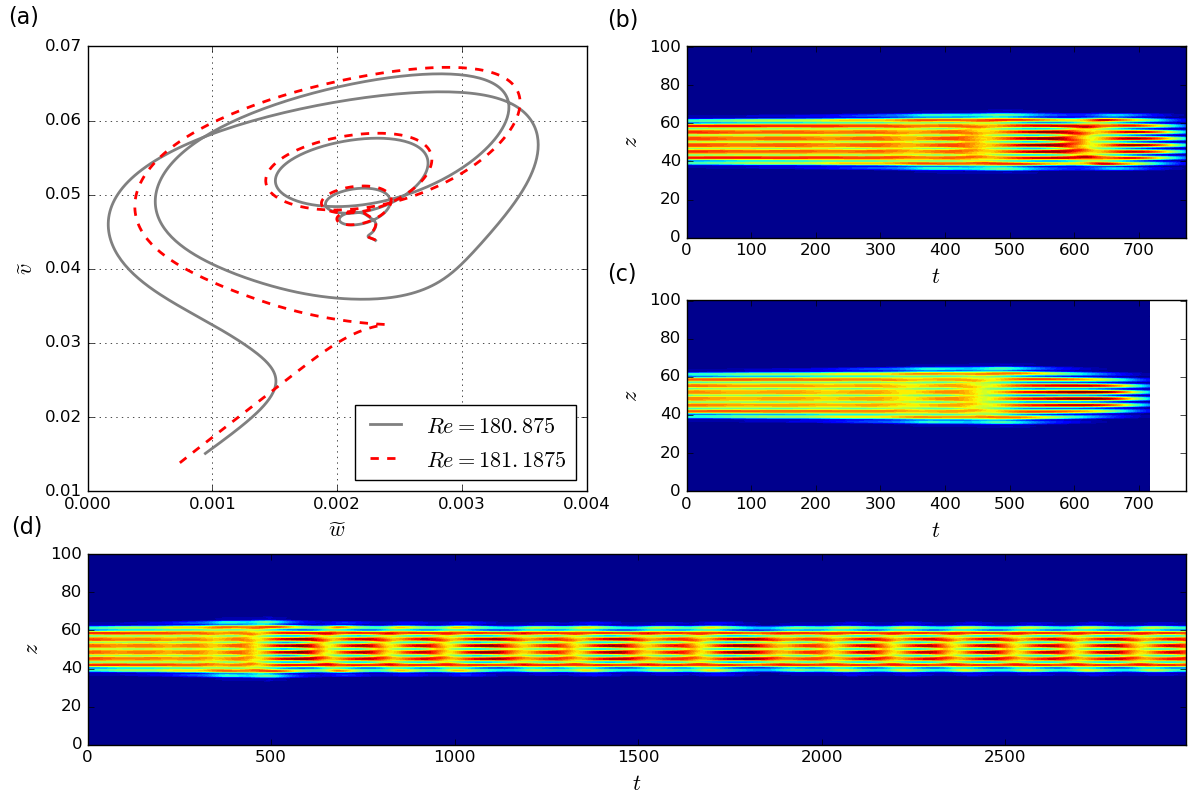}
    \end{center}
    \caption{(a) Trajectories emanating from S5 in R1 and represented through the L2-norms of $v$ and $w$ in the rotated frame of reference ($\widetilde w = ||w|| \cos 70^{\circ} - ||v|| \sin 70^{\circ}$ and $\widetilde v = ||w|| \sin 70^{\circ} + ||v|| \cos 70^{\circ}$) for Reynolds numbers chosen right before ($\Rey = 180.875$) and after ($\Rey = 181.1875$) the peak at $\Rey \approx 181.0625$. Corresponding time-evolution of the averaged kinetic energy $E_{xy}$ before the peak (b) and after the peak (c). (d) Time-evolution of $E_{xy}$ at the peak.}
    \label{fig:r1_crossing_peak}
\end{figure}

\begin{figure}
    \begin{center}
    \includegraphics[width=1\textwidth]{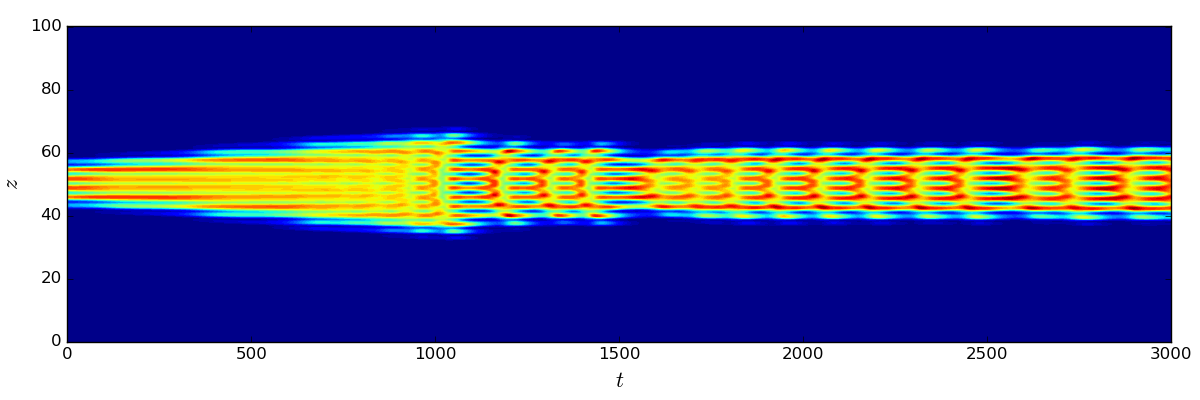}
    \end{center}
    \caption{Heteroclinic connection between the lower branch of EQ and PO5 illustrated by the spatiotemporal evolution of $E_{xy}$ for the exact equilibrium found at the lower branch of EQ at $\Rey = 186.17105$ and perturbed along its most unstable eigendirection.}
    \label{fig:eq_low_po5_heteroconn}
\end{figure}

The diverging relaminarisation times at the peaks are associated with the presence of a periodic orbit.
This phenomenon is shown in figure \ref{fig:r1_crossing_peak}.
On the left of the peak located at $Re \approx 181$, the flow dynamics describes $5$ oscillations which are clearly visible with suitably chosen variables such as the L2-norms of $v$ and $w$ in the appropriately rotated frame of reference.
These oscillations correspond to the non-monotonic departure from the location of the initial condition until the basin of attraction of the laminar solution is reached (panel (b) of figure \ref{fig:r1_crossing_peak}).
As the Reynolds number is increased past the threshold value for the peak, one such oscillation is lost due to the influence of the stable manifold of the laminar solution at shorter time (see the differences between the dashed red and the gray trajectories around $\widetilde v = 0.035$ and $\widetilde w = 0.0025$).
As a consequence, it only takes $4$ oscillations for the flow to decay on the right of this peak, as shown in panel (c).
At the peak (panel (d)), the flow follows the stable manifold of the periodic orbit PO5 whose neighbourhood is reached after about $500$ units of time and where the dynamics shadows that described in Section 4.1. 
This approach to an unstable periodic orbit suggests the existence of heteroclinic connections between steady localized states and periodic orbits in plane Couette flow.
Figure \ref{fig:eq_low_po5_heteroconn} shows such a connection between the lower branch of EQ and PO5 at $\Rey \approx 186.17105$ which was obtained by perturbing along the most unstable eigendirection of the lower EQ branch state and time-integrating.
After the initial sequence of roll nucleation events, the oscillations become a prominent feature of the flow which readjusts and converges to PO5.
 
The behaviour observed above extends to all the peaks in the region R1: every time a peak is crossed in the direction of decreasing values of the Reynolds number, the flow undergoes one more oscillation before relaminarising.
This, together with the peak accumulation at $Re_s$, accounts for the gradual increase of the relaminarisation time.
Conversely, as $\Rey$ in increased past the rightmost peak of R1, the flow only displays two oscillations before eventual decay.

The other initial conditions led to qualitatively similar results with the difference that the peaks obtained in R1 for Si, i$>5$, are not associated with diverging relaminarisation times. 
The right panel of figure \ref{fig:r1_s5_and_s7_relam_time} shows the behaviour of the relaminarisation time in the region R1 for S7 and exemplifies the smooth behaviour of the relaminarisation times at the peaks. 
These peaks correspond to the approach of a periodic orbit constituted of $7$ rolls and analogous to PO5 in dynamics.
The periodic orbit PO5 has only one unstable eigenmode for most of the Reynolds numbers within R1 and, thus, changing the Reynolds number acts as a projection shift of the initial condition onto the unstable manifold of PO5.
At the peak, S5 has exactly no projection onto the unstable manifold of PO5, allowing the relaminarisation time to approach infinity as the trajectory converges to PO5.
Unlike PO5, PO7 has more than one unstable eigendirection and, as a consequence, the projection of S7 onto the unstable manifold of PO7 does not go to zero by simply changing the Reynolds number and the relaminarisation time never approaches infinity.
A similar reasoning holds for the peaks of initial conditions wider than S5.

\begin{figure}
    \begin{center}
    \includegraphics[width=1\textwidth]{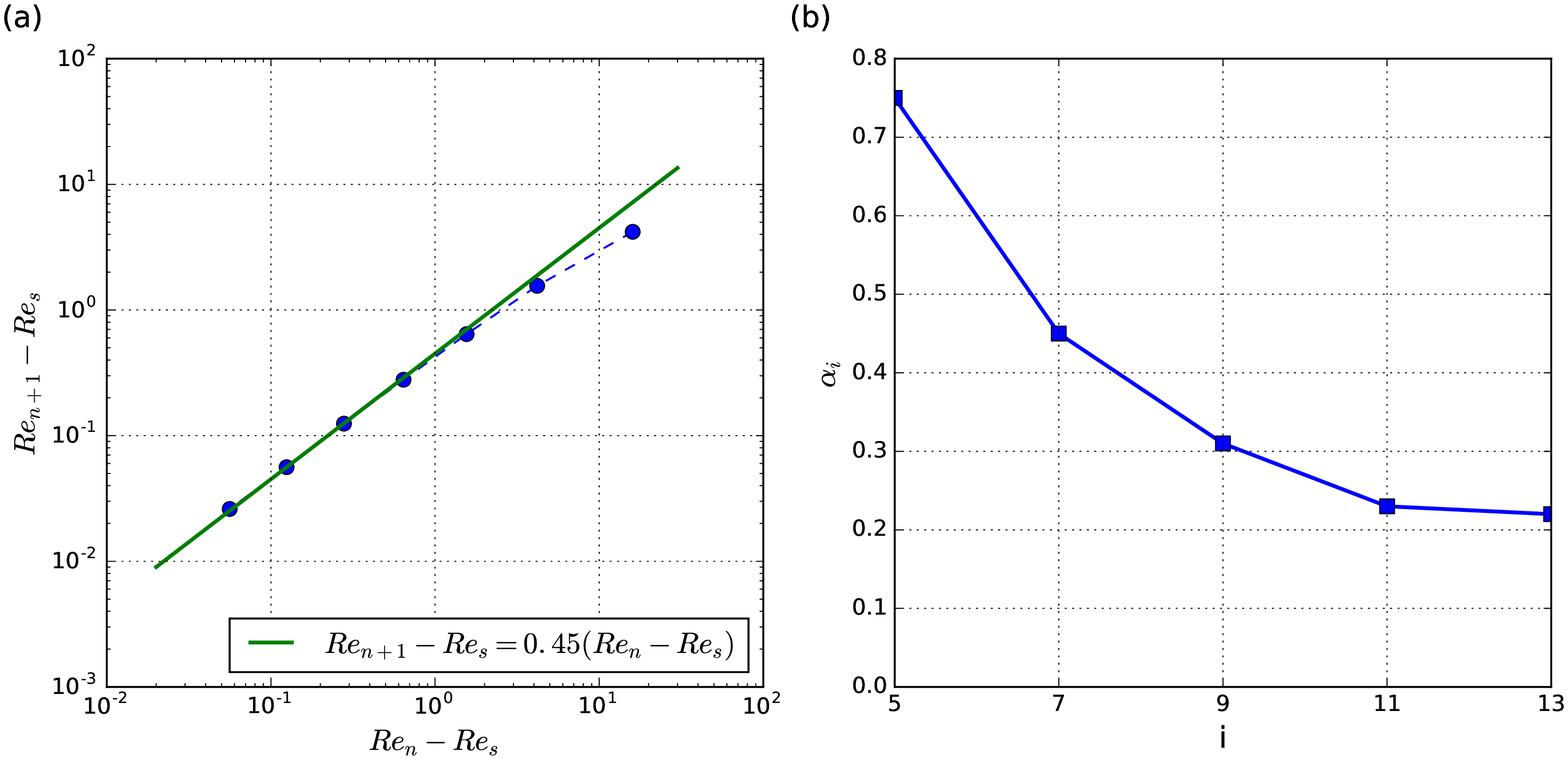}
    \end{center}
    \caption{(a) Peak location for S7 (circles) and the corresponding geometric law $\Rey_{n+1} - \Rey_s = 0.45 \left(\Rey_{n} - \Rey_s \right)$ (solid line). (b) Convergence rates $\alpha_i$ for the five most localized initial conditions. }
    \label{fig:r1_peaks_convergence}
\end{figure}

Peak accumulation is a generic feature of R1 and is characterised by a geometrical convergence law:
\begin{equation}
\label{eq:r1_geom_law}
\Rey_{n + 1} - \Rey_s(\textrm{i}) = \alpha_i \left[ \Rey_{n} - \Rey_s(\textrm{i}) \right],
\end{equation}
where $\Rey_n$ is the Reynolds number corresponding to the $n$-th peak ($n$ is counted from right to left) and $\alpha_i$ the convergence coefficient that is a function of the initial condition Si. 
Figure \ref{fig:r1_peaks_convergence}(a) exemplifies this on S7 and shows very good accuracy between the data and law (\ref{eq:r1_geom_law}) for $\alpha_7 = 0.45$.
The same law applies successfully to other initial conditions, albeit with different convergence ratios $\alpha_i$. 
The corresponding convergence ratios are shown for S5, S7, S9, S11 and S13 in figure \ref{fig:r1_peaks_convergence}(b) and indicate a decreasing trend as the number of rolls i increases, potentially converging to $\alpha_{\infty} \approx 0.2$.
This signifies that as the initial condition increases in size, the accumulation of peaks becomes faster and faster and the width of the region R1 decreases, a fact already observed in figure \ref{fig:relam_180_320} and sketched in figure \ref{fig:regions_schema}. 

The overall decreasing trend of the relaminarisation time with the Reynolds number can be quantified by a difference between two successive local minima of $t_{relam}$.
This quantity, which we denote $\beta$, only has a weak dependence on the initial condition so we will use its value for S5 regardless of the initial condition: $\beta \approx 155$.
\begin{figure}
    \begin{center}
    \includegraphics[width=1\textwidth]{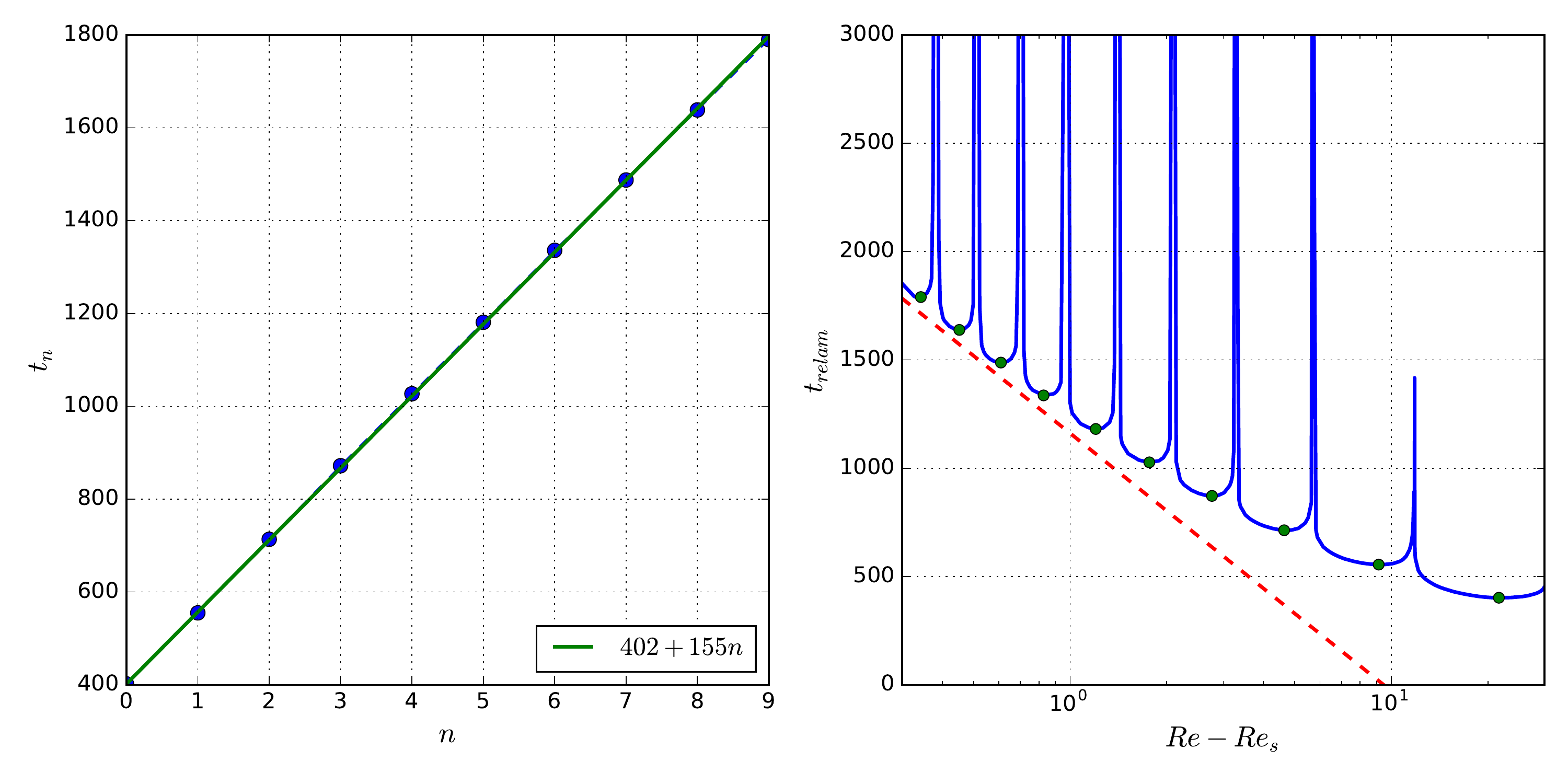}
    \end{center}
    \caption{The left plot shows relaminarisation times against the peak number $n$ at the local minima of the curve in figure \ref{fig:r1_s5_and_s7_relam_time} for S5 (full circles) and an approximation law (solid line). The right plot shows the relaminarisation time for S5 (solid line) together with the first order approximation law for its local minima (circles): equation (\ref{eq:r1_local_minima_law}) (dashed line) for $t_0 = 402$, $\beta = 155$, $\alpha_i = \alpha_5 = 0.74$, $Re_s = Re_s(5)=175.375$ and $Re_0 = 180.375$.}
    \label{fig:r1_local_minima_laws}
\end{figure}
We can thus infer the following approximation for the relaminarisation time at the local minima: $t_{n} = t_{0} + \beta n$, where $t_{n}$ is the value of the local minimum of $t_{relam}$ located to the left of the peak at $\Rey_n$ and $t_{0}$ is the relaminarisation time observed at the local minimum located to the right of the first (highest $Re$) peak.
In order to obtain a relationship between $t_n$ and $\Rey$, we approximate the location of the local minimum $\Rey_{n}'$ to be exactly between the left and right peaks: $\Rey_{n}' = \frac{1}{2}\left(\Rey_{n+1} + \Rey_{n}\right)$. 
Since the solution of equation (\ref{eq:r1_geom_law}) is: $\Rey_n - \Rey_{s} = \left[\Rey_0 - \Rey_{s}\right] \alpha_i^n$, we obtain:
\begin{equation}
\label{eq:r1_local_minima_law}
t_{n} (\Rey_n') = \frac{\beta}{\ln{\alpha_i}} \ln \displaystyle\frac{2 \left(\Rey_{n}' - \Rey_{s} \right)}{\left(\Rey_0 - \Rey_{s} \right) \left(1 + \alpha_i \right)} + t_{0}.
\end{equation}
Extending this law to continuous values of the Reynolds number, we can observe that, close enough to $\Rey_s$, the envelope of the relaminarisation time grows with rate: $dt_{relam}/dRe = O((\ln \alpha_i (Re - Re_s))^{-1})$.
The right plot in figure \ref{fig:r1_local_minima_laws} confirms that this law is a good first order approximation to the relaminarisation time for S5 close to the saddle-node.

\subsection{Onset of chaotic transients}
\label{rr2}

Further from the snaking, our initial conditions display intermittent chaotic transients.
This non-trivial dynamics is partitioned in parameter space into three regions separated by the plateaux described previously.
The first such region is R2 and the solutions, although displaying weak sensitivity to the parameter value, do not typically have relaminarisation times larger than $t_{relam} = 2000$.
Long-lasting chaotic transients can be found in region R3 between $Re = 245$ and $Re = 252$, with relaminarisation times up to $t_{relam} = O(10^4)$ while displaying much more sensitivity to the parameter value.
For $Re > 290$ we enter region R4 where only rare and isolated initial conditions relaminarise within the time periods of interest here.

\subsubsection{Splitting}

\begin{figure}
    \begin{center}
    \includegraphics[width=1\textwidth]{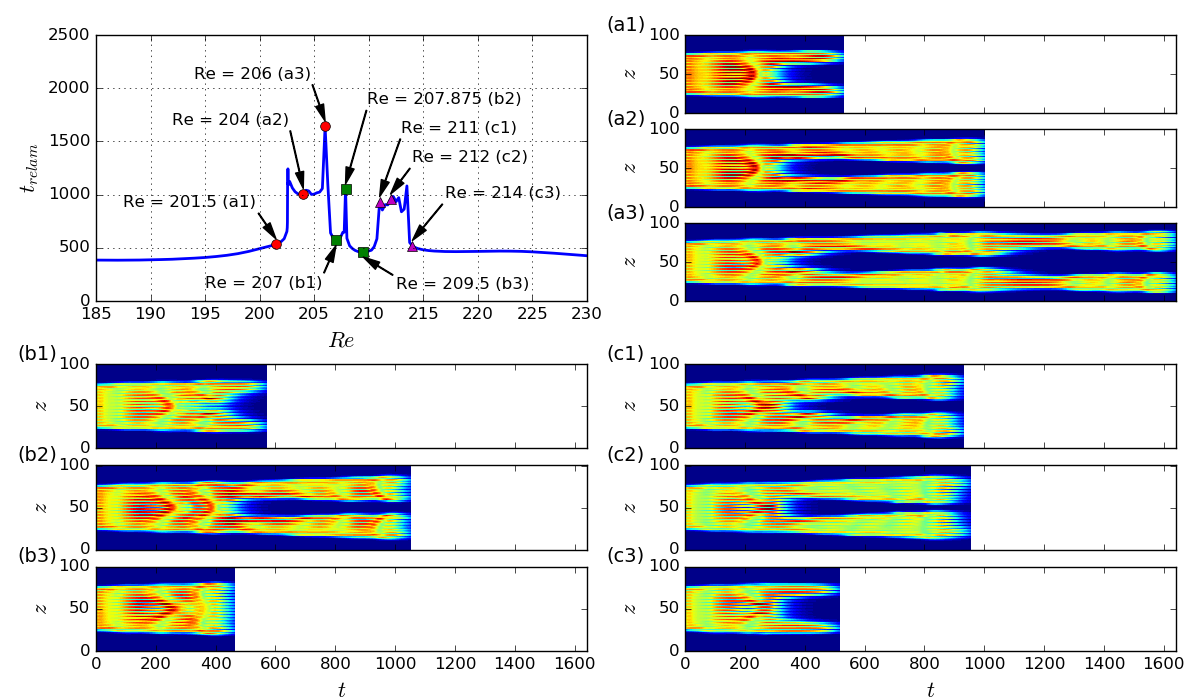}
    \end{center}
    \caption{The upper left plot shows relaminarisation times in R2 for S13. Each of the other nine panels contains a space-time plot of the $xy$-averaged kinetic energy $E_{xy}$ corresponding to simulations grouped into three sets showing (a) the narrowing of the gap between the two pulses, (b) simulations with an extra oscillation of the original pattern and (c) the widening of the gap between the two pulses.
The corresponding Reynolds numbers are located on the upper left plot.}
    \label{fig:r2_splitting}
\end{figure}

The boundaries of the chaotic regions are identified with the splitting of the initial state into two pulses which then undergo oscillations before decaying.
This is best illustrated for the simplest of these regions: R2, which exists for initial conditions wider than S8.
Region R2 is small and close to R3 for the narrower initial conditions (see figure \ref{fig:relam_180_320}).
It moves to smaller Reynolds numbers and expands for wider initial conditions.
Figure \ref{fig:r2_splitting} shows the relaminarisation time within R2 for S13 and space-time plots of $E_{xy}$ for characteristic simulations in R2.
At the beginning of R2 ($\Rey \approx 202.5$), the central part of the localized pattern undergoes decay at $t \approx 300$, forming two dormant spots that survive until $t\approx 550$, as shown in figure \ref{fig:r2_splitting}(a1).
As the Reynolds number is increased, these spots survive for a longer time and become active, undergoing such oscillations as those described in Section \ref{cyclic} and resulting in relaminarisation times of the order of $10^3$ as shown in figure \ref{fig:r2_splitting}(a2).
As the Reynolds number is increased within this region, the number of core rolls that decay during the splitting event decreases until it is temporarily suppressed (see the evolution of the pattern at $t \approx 350$ in figure \ref{fig:r2_splitting}(a1, a2, a3, b1).
This does not rule out the splitting event completely but rather postpones it: the initial pattern can be observed to oscillate twice before splitting until $\Rey \approx 206.5$, but then to oscillate three times before splitting (see panel (b2) for instance).
This behaviour is observed until $Re \approx 211$ where the evolution mirrors that for $Re < 206.5$: splitting occurs after two oscillations and manifests itself by the decay of the two central rolls, the number of these decaying rolls is increasing as $Re$ increases until the edge of R2 (see panels (c1, c2, c3)).

The region R2 is structured as described above until S14 where the gain of one oscillation of the initial pattern is obtained via a sharp peak similar to the peaks observed in R1, instead of an extended region where the inner dynamics change.
We attribute this to the fact that the Reynolds number is too low to sustain spot dynamics.
The left boundary of R2 thus becomes less well-defined and does not approach $\Rey_s(\textrm{i})$ as fast for Si, i$\geq 14$, as for i$<14$ as i increases.

\begin{figure}
    \begin{center}
    \includegraphics[width=1\textwidth]{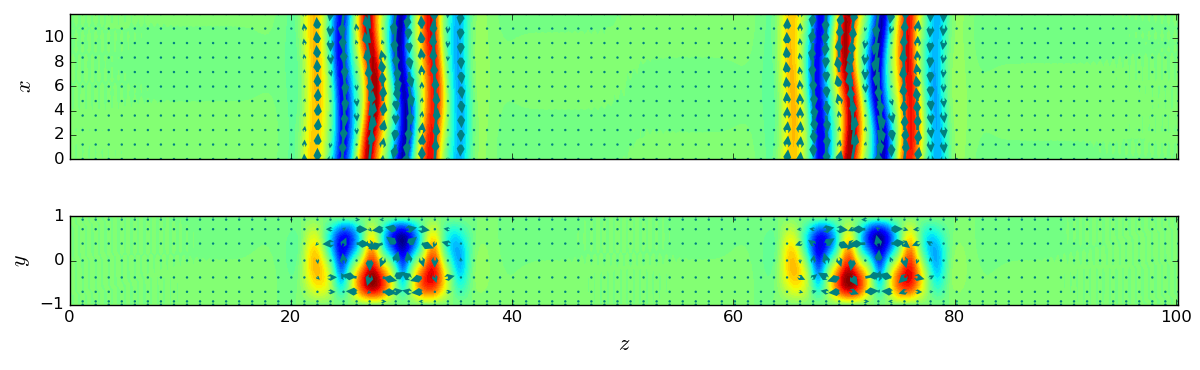}
    \end{center}
    \caption{Example of two-pulse localized equilibrium converged by a Newton--Krylov hookstep search. The initial guess for the search was obtained from the snapshot of the time-integration of S13 at $\Rey = 202.546875$ at time $t = 600$ where the spots formed by the splitting are well separated by the laminar flow.}
    \label{fig:s13_r2_left_boundary_exact_state}
\end{figure}

An interesting feature of the splitting event that characterises R2 is that the spanwise width of the spots immediately after splitting at the boundary of R2 is the same for all initial solutions and coincides with that of the lower part of the branch EQ: the resulting structures are constituted of 3 rolls.
This indicates that the boundaries of R2 correspond to the stable manifold of a fixed point that is a bound state of localized structures similar to the lower part of the branch EQ that are unevenly spaced out. 
This was confirmed by the successful convergence of several solutions taken right after splitting to the above steady state via a Newton--Krylov search. 
Figure \ref{fig:s13_r2_left_boundary_exact_state} exemplifies one such steady two-pulse state converged from the flow snapshot taken from the time-integration of S13 at $\Rey = 202.546875$.
Convergence from other initial conditions can be achieved and would yield a similar state but with different inner/outer spacing.

\subsubsection{Onset of long-lived chaotic transients}
\label{rr3}

\begin{figure}
    \begin{center}
    \includegraphics[width=1\textwidth]{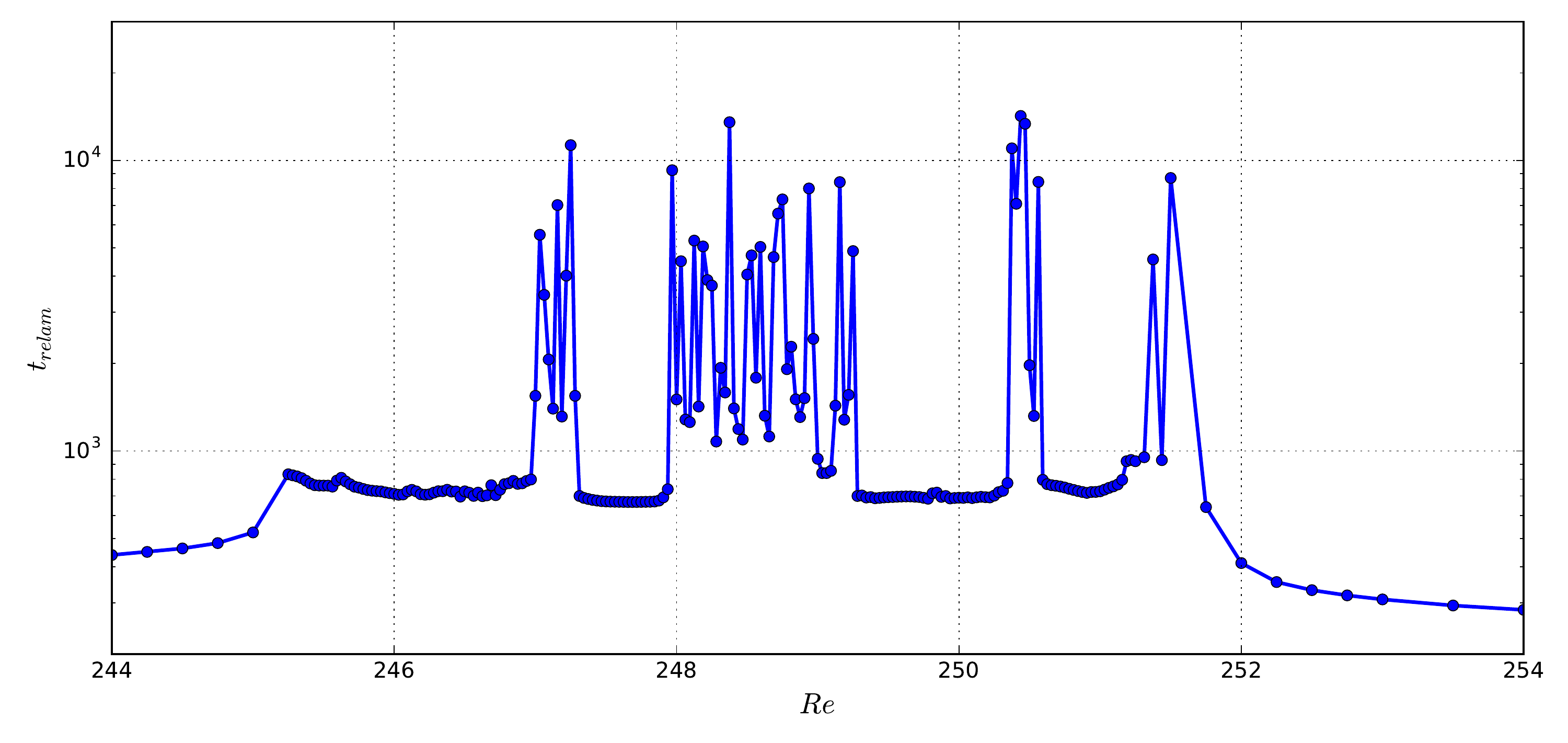}
    \end{center}
    \caption{Relaminarisation time $t_{relam}$ in R3 as a function of the Reynolds number $Re$ for initial condition S15. These results were obtained for Reynolds number increments of $0.03125$.}
    \label{fig:r3_s15_relam_time.eps}
\end{figure}

\begin{figure}
    \begin{center}
    \includegraphics[width=1\textwidth]{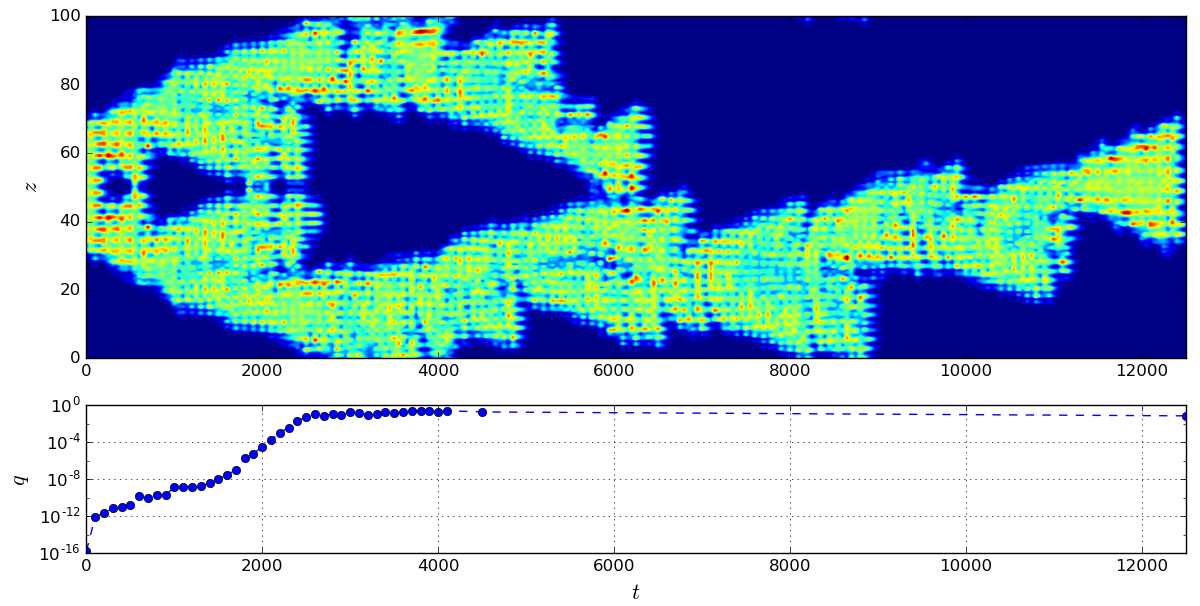}
    \end{center}
    \caption{Space-time plot of the $xy$-averaged kinetic energy $E_{xy}$ for initial condition S9 at $\Rey = 248.75$ in R3 (top) and time-evolution of the reflection symmetry indicator $q$ as explicited in the text (bottom).}
    \label{fig:r3_chaotic_transient}
\end{figure}

The first occurrence of long chaotic transients is found in region R3 which exists for all initial conditions wider than S7 and spans $245 < \Rey < 252$ with little dependence on the initial condition.
In a similar way to R2, its beginning and end are characterised by the splitting of the initial pattern into two spots.
Figure \ref{fig:r3_s15_relam_time.eps} shows the dependence of the relaminarisation time on the Reynolds number for initial condition S15 between $Re = 244$ and $Re = 254$.
For this initial condition, R3 starts at $\Rey \approx 245$ and finishes at $\Rey \approx 251.5$.
Simulations within R3 typically display $700 < t_{relam} < 850$ with most values around $750$ and are only weakly sensitive to the value of the Reynolds number.
Exceptions to this are observed in the form of parameter windows where the relaminarisation time takes much larger values and where the flow becomes strongly sensitive to the Reynolds number.
These windows have various sizes and different initial conditions produce a different number of such windows within R3: e.g. S11 and S15 have 4 such windows each while S9 only has 2.
Inside these windows, the sensitivity to the Reynolds number allows the presence of extremely long chaotic transients, the longest of those we obtained was for S9 at $Re = 251.375$ where $t_{relam} \approx 21320$.
We show in figure \ref{fig:r3_chaotic_transient} such a long simulation for S9, taken at $\Rey = 248.5$ and yielding $t_{relam} \approx 12505$.
This simulation starts in the usual way: the initial pattern splits into two spots at around $t = 200$ which then oscillate and grow spatially by nucleating rolls on either side.
This process continues until about $t = 550$ where both spots collide.
Shortly after, the central part of the structure collapses in an event that typically leads to the complete relaminarisation of the flow.
In this specific instance, the ``outer'' parts of the structure manage to survive and continue their independent evolution.
The flow dynamics that ensues is long-lived and likely similar to that observed in percolation dynamics \citep{Pomeau1986, Barkley16, Lemoult:2016_directed_percolation_in_couette_flow}.
Chaotic transients have also been observed at such low Reynolds numbers by \cite{Schmiegel2000} and \cite{Barkley2005} using different methods.

\begin{figure}
    \begin{center}
    \includegraphics[width=1\textwidth]{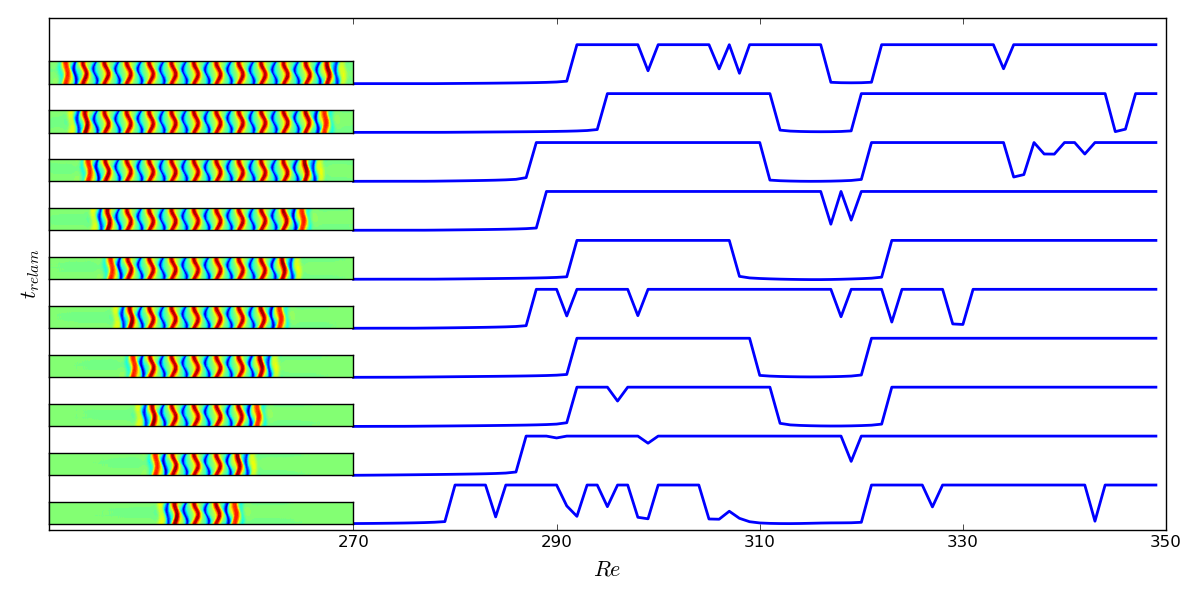}
    \end{center}
    \caption{Relaminarisation times in R4 for the right saddle-nodes states of EQ time-integrated for $\Rey \in [270; 350]$. Relaminarisation times were cut off at $t_{relam} = 2000$ to avoid overlapping and save computation time. As a consequence, the large plateaux where $t_{relam} = 2000$ correspond to long-lasting chaotic transients that extend beyond $t=2000$.}
    \label{fig:relam_270_350}
\end{figure}

\begin{figure}
    \begin{center}
    \includegraphics[width=1\textwidth]{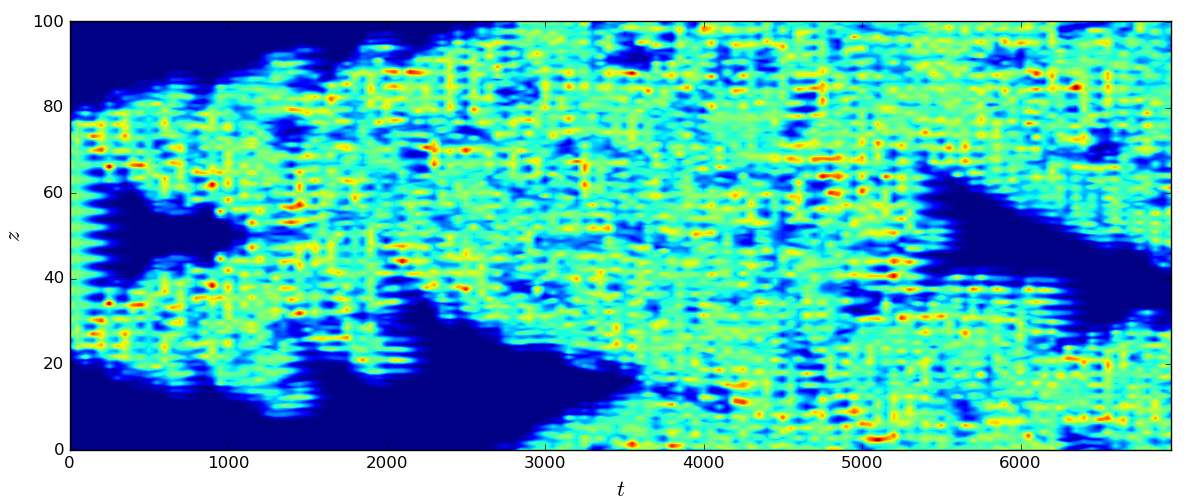}
    \end{center}
    \caption{Space-time plot of the $xy$-averaged kinetic energy $E_{xy}$ for initial condition S13 at $\Rey = 300$ in R4. Relaminarisation is not shown.}
    \label{fig:r4_chaotic_transient}
\end{figure}

Lastly, at $Re \approx 290$, a brutal transition takes place: immediately below the threshold value, all simulations relaminarise in about $300$ units of time whereas we only observed rare cases where relaminarisation occurs in less than $2000$ time units above the threshold as can be seen from figure \ref{fig:relam_270_350}.
This transitional region, R4, exhibits front propagation and long timescales and suggests transition to turbulence, as it is understood in the statistical sense \citep{Avila:2011_onset_of_turbulence,Shi:2013_scale_invariance_at_onset_in_pcf,Barkley16}.
In spite of the long timescales, the dynamics in R4 is different from that observed in R3, where we observe a competition between the growth of the structure via front propagation and the sudden decay of roll clusters (see figure \ref{fig:r3_chaotic_transient}).
Here, front propagation overwhelms cluster decay and leads to domain-filling states.
Figure \ref{fig:r4_chaotic_transient} shows a typical R4 simulation, observed for S13 at $Re = 300$.
The presence of relatively large number of long-lived simulations in R4 allows the quantification of the front speed. 
We did not find any significant difference in front speed between initial conditions.
Additionally, there was no apparent variation in the front speed in the interval $\Rey \in [280; 350]$; the average front speed value is $\langle c \rangle \approx 0.02$.
These results compare very well with \citet{Duguet:2011_stochastic_and_deterministic_front}.

Since the equilibrium initial conditions under consideration are all reflection symmetric, it is important to understand how the flow departs from the symmetry subspace.
We quantify the deviation from the reflection symmetry by:
\begin{equation}
q(t) = \left(\frac{1}{\Gamma_x \Gamma_y \Gamma_z} \int \int \int_{\Omega} \left[u(x, y, z, t) + u(-x, -y, -z, t)\right]^2 dx dy dz \right)^{\frac{1}{2}}.
\end{equation}
A typical time-evolution for this quantity is shown in the bottom panel of figure \ref{fig:r3_chaotic_transient}. 
The deviation from the reflection symmetry grows until it reaches $O(1)$ values at $t \approx 2500$.
Similar evolutions of $q$ are observed in other simulations and imply that asymmetry becomes important only for $t \gtrsim 2000$.
All shorter simulations can be treated as fully symmetric ones.
This was confirmed by the recalculation of one of the relaminarisation curves from figure \ref{fig:relam_180_320} with imposed reflection symmetry.
The only simulations exhibiting significant asymmetry correspond to chaotic transients with long enough lifetime which occur only in R3 and R4. 

\section{Discussion}

In this paper, we have studied plane Couette flow in a periodic domain with large spanwise and short streamwise periods.
We used spatially localized states found on snaking branches as initial conditions and investigated their transitional dynamics beyond the snaking.
We varied two parameters: the size of the localized pattern constituting the initial condition and the Reynolds number.
By tracking the relaminarisation time $t_{relam}$ for each simulation, we identified plateaux in parameter space where the initial condition relaminarises quickly and a hierarchy of regions where it evolves into a more or less long-lived chaotic transient.
We find that the dynamics in the vicinity of the snaking is controlled by an oscillatory instability whose growth leads to the crossing of a manifold leading to relaminarisation.
The number of oscillations undergone by the flow before crossing this manifold is a function of the Reynolds number and the width of the initial localized pattern.
It is well described by our ad-hoc model.
Farther away from the snaking, one encounters various regions of increasing complexity initiated by the splitting of the original pattern into two spots and characterised by increasingly long relaminarisation times and an increasingly high sensitivity to the location in parameter space (sensitivity to the Reynolds number but also to the width of the patterns and to other small perturbations).

\begin{figure}
    \begin{center}
    \includegraphics[width=1\textwidth]{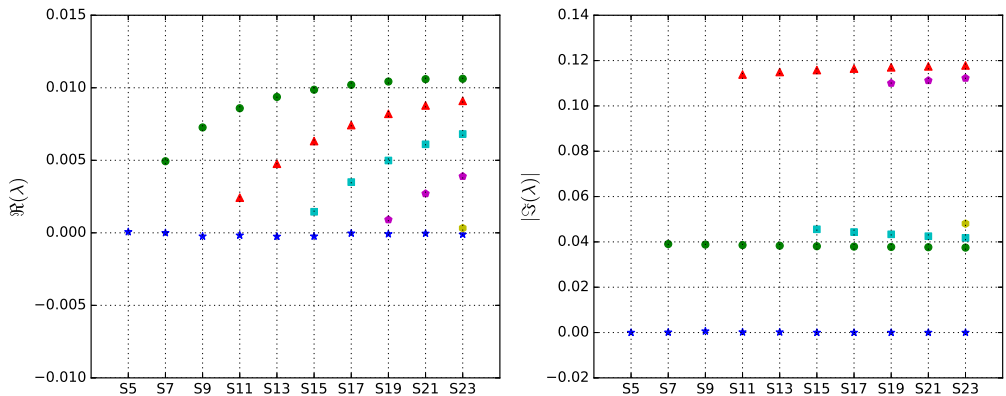}
    \end{center}
    \caption{Unstable eigenvalues $\lambda$ of the saddle-node states of the branch EQ. The left (resp. right) panel show their real (resp. absolute imaginary) part. Blue stars denote the eigenvalue associated with the depinning mode. Green circles denote the eigenvalue associated with 1-pulse oscillatory instabilities that trigger amplitude oscillations of the whole localized structure. Red triangles, cyan squares, purple pentagons and yellow hexagon denote the eigenvalues associated with oscillatory instabilities triggering 2-, 3-, 4- and 5-pulse oscillations.}
    \label{fig:saddle_nodes_stability}
\end{figure}

Contrary to \citet{Duguet:2011_stochastic_and_deterministic_front}, we did not find any signature of depinning, a behaviour associated with snaking and in which the localized state grows by successive roll nucleations and with a frequency that depends on the distance to the snaking.
Although this instability occurs in other systems \citep{Coullet00,Saarloos03,Knobloch:2015_spatial_localization}, it is not expected in all systems that exhibit snaking; for example, three-dimensional doubly diffusive convection displays an instability whose growth rate is much larger then that of depinning responsible for the decay of the localized patterns before any depinning event could take place \citep{Beaume18}.
The spatially localized snaking states we obtained in this paper are also unstable to additional oscillatory instabilities which dominate the dynamics: they lead to the direct relaminarisation of the flow at low values of the Reynolds number and to splitting then relaminarisation at larger $Re$.
Figure \ref{fig:saddle_nodes_stability} shows the unstable eigenvalues associated with the right saddle-node states of the branch EQ. 
The depinning mode is marginal at the saddle-node but we observe the presence of unstable oscillatory eigenmodes for all initial conditions of width equal or greater to that of S7.
These instabilities grow on $O(100)$ timescales and, as a result, prevent depinning from being observed.
Another complication arises in plane Couette flow: snaking is sensitive to the imposed streamwise period of the flow \citep{Gibson:2016_homoclinic_snaking}. 
Our choice ($L_x = 4\pi$) yields snaking in $170 < Re < 175$ while \citet{Duguet:2011_stochastic_and_deterministic_front} chose $L_x = 10.417$ and observe snaking in $207.4 < Re < 213.2$.
This difference also affects the stability of the localized states: in the latter case, non-depinning modes are associated with longer timescales and allow for depinning to be observed.

Understanding the splitting instability is critical to gain further knowledge on long-lived chaotic transients.
This instability is the consequence of the crossing of the stable manifold of a two-pulse exact (either equilibrium or travelling wave) solution.
One side of this stable manifold leads to relaminarisation while the other leads to the activation of the newly formed spots.
The most interesting of these two-pulse states is the one that has the smallest number of rolls, corresponding to the onset for the region dynamics.
One step in understanding this instability further would be to use these two-pulse states as a proxy for initial conditions in a relaminarisation study similar to that described here.

The cyclic dynamics described in Section \ref{cyclic} supports most of the low Reynolds number dynamics observed in this paper.
The oscillations described here likely relate to those found in \cite{Barkley2005} occurring at similar $Re$ and with similar oscillation period.
This cyclic dynamics is reminiscent of the self-sustaining process (SSP) \citep{Waleffe:1997_on_ssp,Waleffe:2009_Exact_coherent_structures} in which a three-step equilibrium loop was found to support the existence of exact solutions in shear flows.
The observed cycles consist in heteroclinic connections between the lower and upper branch Nagata solutions and the mechanisms by which they are sustained seem more complex than the SSP.
As such, the oscillatory dynamics observed here looks similar to the EQ1--EQ2 heteroclinic connection found in \cite{Halcrow:2009_Heteroclinic_connections}.
The lower branch Nagata solution is known to be an SSP state \citep{Wang:2007_lower_branch}, but the upper branch one has been known to not have the same scaling and thus to not obey the same balance.
This cyclic process can then be thought of as related to the SSP and the regeneration cycle \citep{Hamilton1995, Kawahara2001} whilst organising the flow dynamics at low values of the Reynolds number and supporting the existence of some periodic orbits.
It may also be relevant to the relative periodic orbits found by \citet{Viswanath:2007_recurrent_motions_within_pcf} (solutions $P_2$ through $P_6$) which seem linked to the recurrent bursting events observed experimentally at higher Reynolds numbers.
Our periodic orbit PO5 might be thought as a localized counterpart of these solutions where drifting in the spanwise direction is prevented by the pinned fronts.

Finally, the maps of relaminarisation times for low $\Rey$ (figure \ref{fig:relam_180_320}) and transitional $\Rey$ (figure \ref{fig:relam_270_350}) together with the information about dynamical properties of the flow can be used as a framework for the design and benchmarking of control strategies aimed at controlling transition to turbulence.
The statistics extracted from the relaminarisation times and the Reynolds numbers of the identified regions may provide one of the possible sources of information for the quantification of the control efficiency.
Owing to the oscillatory nature of the dynamics observed, it is natural to propose control strategies that interact with the flow frequencies exploiting resonance mechanisms.
Given the spanwise localization of our initial conditions, the aforementioned results will remain valid for domains of sufficiently large spanwise extent and of streamwise periodicity $4 \pi$.

\section{Acknowledgments}

We are grateful to Prof. John Gibson for assistance with the computations. AP acknowledges EPSRC for supporting him through a Doctoral Training Partnership Studentship. This work was undertaken on ARC3, part of the High Performance Computing facilities at the University of Leeds, UK.

\bibliographystyle{jfm}
\bibliography{dynamics_of_localised_states_jfm}

\end{document}